\documentclass[fleqn,usenatbib]{mnras}

\usepackage{cancel}
\usepackage[dvipsnames]{xcolor}
\usepackage{ae,aecompl}
\usepackage{natbib}
\usepackage{graphicx}
\usepackage{array}
\usepackage{amsmath}
\usepackage{float}
\usepackage{amssymb}
\usepackage{subcaption}
\usepackage{newtxtext,newtxmath}
\usepackage[T1]{fontenc}
\usepackage{multirow}
\usepackage{cleveref}

\crefformat{section}{\S#2#1#3} 
\crefformat{subsection}{\S#2#1#3}
\crefformat{subsubsection}{\S#2#1#3}

\newcommand*{\mr}{\mathrm}
\newcommand*{\msun}{\mathrm{M}_\odot}

\title{Detecting Low-Mass Perturbers in Cluster Lenses using Curved Arc Bases}

\author[A. Ç. Şengül, S. Birrer, P. Natarajan, C. Dvorkin]{
Atınç Çağan Şengül$^{1}$\thanks{sengul@g.harvard.edu},
Simon Birrer$^{2}$, Priyamvada Natarajan$^{3,4,5}$, and Cora Dvorkin$^{1}$\\
$^{1}$Harvard University, Department of Physics, Cambridge, Massachusetts, 02138, U.S.A. \\
$^{2}$Stony Brook University, Department of Physics and Astronomy, Stony Brook, New York, 11794, U.S.A.
\\
$^{3}$Department of Astronomy, Yale University, 52 Hillhouse Avenue, New Haven, CT 06520, USA
\\
$^{4}$Department of Physics, Yale University, P.O. Box 208121, New Haven, CT 06520, USA
\\
$^{5}$Black Hole Initiative, Harvard University, 20 Garden Street, Cambridge, MA 02138, USA}

\pubyear{2023}
\hypersetup{final}
\begin{document}
\label{firstpage}
\pagerange{\pageref{firstpage}--\pageref{lastpage}}
\maketitle

%%%%%%%%%%%%%%%%%%%%%%%%%%%%%%%%%%%%%%%%%%%%%
\begin{abstract}
Strong gravitationally lensed arcs produced by galaxy clusters have been observationally detected for several decades now. These strong lensing constraints provided high-fidelity mass models for cluster lenses that include substructure down to $10^{9-10}\,\mathrm{M}_\odot$. Optimizing lens models, where the cluster mass distribution is modeled by a smooth component and subhalos associated with the locations of individual cluster galaxies, has enabled deriving the subhalo mass function, providing important constraints on the nature and granularity of dark matter. In this work, we explore and present a novel method to detect and measure individual perturbers (subhalos, line-of-sight halos, and wandering supermassive black holes) by exploiting their proximity to highly distorted lensed arcs in galaxy clusters, and by modeling the local lensing distortions with curved arc bases. This method offers the possibility of detecting individual low-mass perturber subhalos in clusters and halos along the line-of-sight down to a mass resolution of $10^8\,\mathrm{M}_\odot$. We quantify our sensitivity to low-mass perturbers ($M\sim 10^{7-9}\,\mathrm{M}_\odot$) in clusters ($M\sim 10^{14-15}\mathrm{M}_\odot$), by creating realistic mock data. Using three lensed images of a background galaxy in the cluster SMACS J0723, taken by the {\it James Webb Space Telescope}, we study the retrieval of the properties of potential perturbers with masses $M=10^{7-9}\,\mathrm{M}_\odot$. From the derived posterior probability distributions for the perturber, we constrain its concentration, redshift, and ellipticity. By allowing us to probe lower-mass substructures, the use of curved arc bases can lead to powerful constraints on the nature of dark matter as discrimination between dark matter models appears on smaller scales.
\end{abstract}
%%%%%%%%%%%%%%%%%%%%%%%%%%%%%%%%%%%%%%%%%%%%%

%%%%%%%%%%%%%%%%%%%%%%%%%%%%%%%%%%%%%%%%%%%%%
%%%%%%%%%%%%%%%%%%%%%%%%%%%%%%%%%%%%%%%%%%%%%

%%%%%%%%%%%%%%%%%%%%%%%%%%%%%%%%%%%%%%%%%%%%%
\section{Introduction}\label{sec:intro}
In the standard model of cosmology ($\Lambda$CDM), dark matter is assumed to be a collisionless particle that interacts with itself and ordinary matter only via gravity (known as ``cold dark matter (CDM)"). $\Lambda$CDM has been successful in explaining structure formation in the Universe when applied to distances larger than $\sim 1$ Mpc, which correspond to masses larger than $10^{11} \msun$ \citep{planck2018,boss_matter_power,des_matter_power,cosmic_shear,lyman_alpha,large_scales}.
Despite these successes, the putative dark matter particle remains elusive and has yet to be detected, either directly via tailored experiments or with indirect detection techniques. This lack of detection of the dark matter particle has spurred the examination of alternate dark matter models beyond CDM. Measuring structure formation at sub-galactic scales is crucial to investigating these various dark matter models, as the discrimination between dark matter models manifests on small-scales. Counter-intuitively, as demonstrated here, we can use one of the largest known structures to make these measurements.

Galaxy clusters are the most massive gravitationally bound objects in the Universe, with masses as high as $10^{15}\msun$. Clusters contain thousands of gravitationally bound member galaxies which vary in luminosity. The path of the light rays from distant background galaxies gets deflected by the mass distribution of massive foreground clusters, which results in the production of multiple highly-distorted and magnified images of the distant background sources. This phenomenon, {\it strong gravitational lensing}, is predicted by General Relativity and has come to provide an essential tool for measuring the detailed mass distribution within clusters, giving a detailed census of dark matter (for a review of cluster lenses, see \cite{cluster_lens_review}). Using the shapes, positions, and brightnesses of the multiply imaged sources, lens modeling algorithms have been developed to reconstruct the detailed mass distribution, predominantly dark matter, in cluster lenses. Various independent approaches to modeling the mass distribution within clusters with gravitational lensing have been shown to produce reliable results \citep{Frontier_Fields} down to $10^{9-10} \msun$ \citep{Natarajan+2017,mass_range}, corresponding to scales of $\sim 10$ kpc. In most of these studies, the locations of cluster galaxies are assumed to signal the presence of their associated dark matter subhalos. And this connection between mass and light permits mapping of the subhalo mass function (SHMF) within clusters \citep{Natarajan_1997}. 

In the study of galaxy-scale lenses, where both the source and the lens are galaxies, the source-light distribution is modeled simultaneously with the lens-mass distribution to fit the values of individual pixels in the data. By utilizing the information contained in the full-resolution imaging, one can probe masses down to $10^8 \msun$ in individual galaxy-galaxy strong lensing systems. A parametric power-law model is found to approximate the mass distribution of a galaxy-sized halo in a cluster. The low model-complexity, combined with a small data vector coming from the smaller size (typically $\sim 1''$) of the lensed images, makes sampling the model parameter posteriors computationally tractable.

Cluster lenses can form multiple images that span $\sim 10''$, resulting in a large number of relevant pixels for which we need to compute the lensing observables, of the order of $\sim 10^6$ pixels (compared to $\sim 10^3$ pixels for galaxy-scale lenses), which makes pixel-level modeling of these images computationally challenging for the purposes of parameter estimation. Moreover, the mass distribution within the cluster is much more complex than that within an individual galactic halo, which demands either a free-form approach or a parametric approach that utilizes the observed light from the cluster members as constraints for the mass distribution. Therefore, for practical reasons, when modeling strong lensing clusters, the data vector consists of the image positions and magnifications of the larger number of available lensed galaxies. However, there is much more information to be mined in the lensed images that are detected in cluster lenses that is under-utilized when the data is reduced just to a summary of image positions and magnifications as is currently done. To answer questions that involve large-scale angular deflections, statistical constraints on smaller scales, and direct comparison with cosmological simulations, current approaches to modeling cluster lensing are adequate. However, in order to push further and probe down to the smallest detectable scales to infer the presence of individual low-mass perturbers, we need a new way of handling the full-resolution image data without compressing the data vector to take full advantage of all the available information.

Measured UV-luminosity functions of nearby clusters, including Coma and Virgo, point to the existence of a large population of extremely faint member galaxies \citep{DePropis_2018}. On the theory side, $\Lambda$CDM predicts the existence of abundant subhalos that are available to host either low-luminosity galaxies or even dwarf galaxies that have baryons that are yet to form a substantial population of stars \citep{Tremmel_2018}. Therefore, the characterization of individual perturbing subhalos that do not currently have associated bright observational counterparts would not be captured in current parametric lens models but are predicted to exist. These offer additional leverage and tests of $\Lambda$CDM and the nature of dark matter. The method used in this work permits the detection of such perturbers in addition to the detection of the possible presence of wandering supermassive black holes (SMBH) with masses ranging from $10^{6-9}\,\msun$ in cluster that are also predicted in state-of-the-art simulations, such as the Romulus-suite \citep{wandering_bh}. 

In this work, we propose a novel approach that involves locally modeling multiply imaged \textit{extended} sources in cluster lenses. By only modeling the angular deflections in the specific regions where the images form, using a method called the \textit{curved arc basis} \citep{curved_arc_theory}, we adapt to the current computational challenges of cluster lens modeling. This paper is the first application of the curved arc basis to real data, which so far has been demonstrated solely on simulated data. Our goal is not to reconstruct the mass distribution of the entire cluster but to detect and measure individual low-mass perturbers, both within the cluster and along the line-of-sight, along a narrow cylinder down to the limit of the high-resolution {\it James Webb Space Telescope} (JWST) imaging in this instance. The computational efficiency of this technique allows us to potentially test different dark matter scenarios on small scales within clusters and along the line-of-sight.

This paper is organized as follows: we first present a brief review of the current constraints on perturbing subhalo populations in cluster lenses in \cref{sec:synopsis} before describing our new methodology in \cref{sec:methods}. Results of the application of the method to JWST data of the cluster lens SMACS 0723 are presented in \cref{sec:smacs}; mock tests demonstrating the power of recovery of individual perturber properties with the method are shown in \cref{sec:JWSTsims}, followed by conclusions and discussion in \cref{sec:conclusion}.

%%%%%%%%%%%%%%%%%%%%%%%%%%%%%%%%%%%%%%%%%%%%%

\section{Synopsis of current constraints on substructure in cluster lenses}\label{sec:synopsis}

Before we dive into our methodology in \cref{sec:methods}, we offer a summary of the current status of substructure studies in the context of modeling cluster lenses. There have been many recent developments and improvements in the modeling precision of the overall mass distribution in clusters that include the combining of strong lensing with additional observational data of stellar kinematics \cite{monna2016a,Bergamini2019}; that use features identified in multiple images to model extended shapes of sources and that utilize the information in the extended source light distributions to improve the overall fidelity of lens models \citep{Pignataro2021,Bergamini2021,Diego2022,Sharon2022}. Our goal in this work is, however, somewhat different, as we focus on specifically mapping individual lower mass substructures with $M\,<\,10^{10}\,\msun$, both in the cluster and along the line of sight that may not necessarily even host a bright stellar component. Interestingly, relevant to our work, some estimates have been made of the properties of more massive individual subhalos in cluster lenses, for instance, \cite{parry2016} have measured the mass distribution of a pair of heavier subhalos with masses $M > 10^{10}\,\msun$ within the galaxy cluster MACS J1115.9+0129 using strong lensing constraints and \cite{monna2017} report the measurement of the velocity dispersion and ellipticities; as well as constraints on the core radius and truncation radius of the three central galaxies in the cluster lens MACS 2129.

Meanwhile, current state-of-the-art cluster mass-reconstruction methods, provide {\bf statistical constraints} on the entire subhalo population inside clusters, as opposed to specific individual sub-halos as they typically deploy scaling relations between mass and light in the modeling \citep{Natarajan+2017,mass_range}. Parametric modeling techniques characterize the mass distribution of cluster lenses as being composed of a smooth dark matter component and the sum of the contributions of a population of perturbing subhalos. Typically, in parametric and hybrid methods, self-similar profile shapes are assumed for the larger scale smooth component and the perturbers in the modeling. Using these priors, the positions and brightnesses of the detected families of multiple images and the measured shear are used to obtain the best-fit mass model that can reproduce the observations. The adoption of empirically motivated scaling relations like the Faber-Jackson law between mass and light serves to reduce the dimensionality of the problem. Usually, the full extended pixellated shapes of the highly distorted arcs are not used, and this additional information is discarded, as their ellipticity and measured length to width ratio are used a proxies. The best-fit mass model can be used to derive the SHMF as the integrated mass within an aperture is obtained for subhalos, i.e., mass enclosed within a characteristic cut radius for the adopted parametric profile used to model the perturbers. With high-resolution {\it Hubble Space Telescope} (HST) and JWST cluster lensing data, one is able to map the SHMF down to masses of $10^{9-10}\,\msun$ and these derived SHMFs are in good agreement with predictions of $\Lambda$CDM simulations. Although a new tension has emerged in the recent years. \cite{meneghetti2020} and \cite{meneghetti2022} has shown an excess of small-scale lensing effects in a number of cluster lenses compared to what is expected from $\Lambda$CDM simulations. 

The amplitude and slope of the SHMF are concrete predictions of $\Lambda$CDM. Current parametric lensing mass models have been designed to test this clear-cut prediction. However, at the present time, for low-mass perturbers with masses $M < 10^{10}\,\msun$, and by virtue of construction, we are insensitive to how the mass is distributed in detail spatially within individual subhalos, i.e., the core radius, ellipticity, and density profile of subhalos cannot be constrained. These techniques provide the integrated mass within an aperture for subhalos. 

The new technique presented in this work permits a more complete characterization of individual perturbing low-mass subhalos, though not of the entire population. With galaxy-galaxy strong lensing (GGSL) events within clusters increasingly starting to be resolved with deeper, higher-resolution JWST and HST data, this improvement in methodology is warranted to mine all the information available in the images. Additionally, with the power to constrain the redshift, our method permits breaking a key degeneracy regarding the location of the subhalo in the lens plane versus along the line-of-sight, a long-standing issue in the modeling of galaxy lenses.

%%%%%%%%%%%%%%%%%%%%%%%%%%%%%%%%%%%%%%%%%
\section{Methodology}\label{sec:methods}

In this section, we discuss how we derive constraints on the angular deflection field near the magnified arcs from imaging data. Our data vector $\mathbf{D}$ consists of the pixels of the cutouts of multiple images of a source galaxy lensed by a cluster (see Fig. \ref{fig:smac_wide}). Our model reconstruction, denoted $\mathbf{M}(\mathbf{q})$, is a function of a set of model parameters represented by the vector $\mathbf{q}$. When a telescope makes an observation, the captured image is the convolution of the surface brightness distribution with the point spread function (PSF), sampled by a pixelated grid, which determines the resolution of the instrument. We denote this operation with the operator $\mathcal{T}$ so that
\begin{equation}\label{eq:telescope}
    \mathbf{M}(\mathbf{q}) = \mathcal{T}\left\{\mathbf{I}(\mathbf{q})\right\},
\end{equation}
where $\mathbf{I}(\mathbf{q})$ is the model reconstruction of the surface brightness distribution. This surface brightness is the result of the gravitational lensing of the source light distribution. Lensing can also be written as an operation or mapping, denoted by the operator $\mathcal{L}(\mathbf{q_\mr{lens}})$, applied on the source light model $S(\mathbf{q_\mr{source}})$ to give the surface brightness reconstruction:
\begin{equation}\label{eq:lensing_op}
    \mathbf{I}(\mathbf{q}) = \mathcal{L}(\mathbf{q_\mr{lens}})\left\{S(\mathbf{q_\mr{source}})\right\}.
\end{equation}
The lens parameters $\mathbf{q_\mr{lens}}$ and source parameters $\mathbf{q_\mr{source}}$ are the components of the full parameter vector $\mathbf{q} = (\mathbf{q_\mr{lens}},\mathbf{q_\mr{source}})$. The lensing operator uses the conservation of surface brightness under lensing, which states that the surface brightness $\mathbf{I}(\vec x)$ at an angular position $\vec x$ on the sky is equal to the surface brightness of the source $\mathbf{I}_S(\vec y)$ at the corresponding angular position $\vec y$ on the source plane. The relation between the two angular positions is given by the lens equation $\vec y = \vec x - \vec \alpha(\vec x)$, where $\vec \alpha$ is the angular deflection. The lensing operator $\mathcal{L}$ is fully determined by the angular deflection field $\vec \alpha$. We will describe our lens model by either explicitly providing the angular deflections as a vector field, or by providing the mathematical form of the mass distribution causing the lensing deflection. Using Eqs. \eqref{eq:telescope} and \eqref{eq:lensing_op} the model reconstruction can be written as:
\begin{equation}\label{eq:telescope2}
    \mathbf{M}(\mathbf{q}) = \mathcal{T}\big\{\mathcal{L}(\mathbf{q_\mr{lens}})\left\{S(\mathbf{q_\mr{source}})\right\}\big\}.
\end{equation}
We use \textsc{lenstronomy} \citep{lenstronomy1,lenstronomy2}, a publicly available gravitational lensing package, to perform the calculations for the model reconstruction in this work.

\begin{figure}
    \centering
    \includegraphics[width=\linewidth]{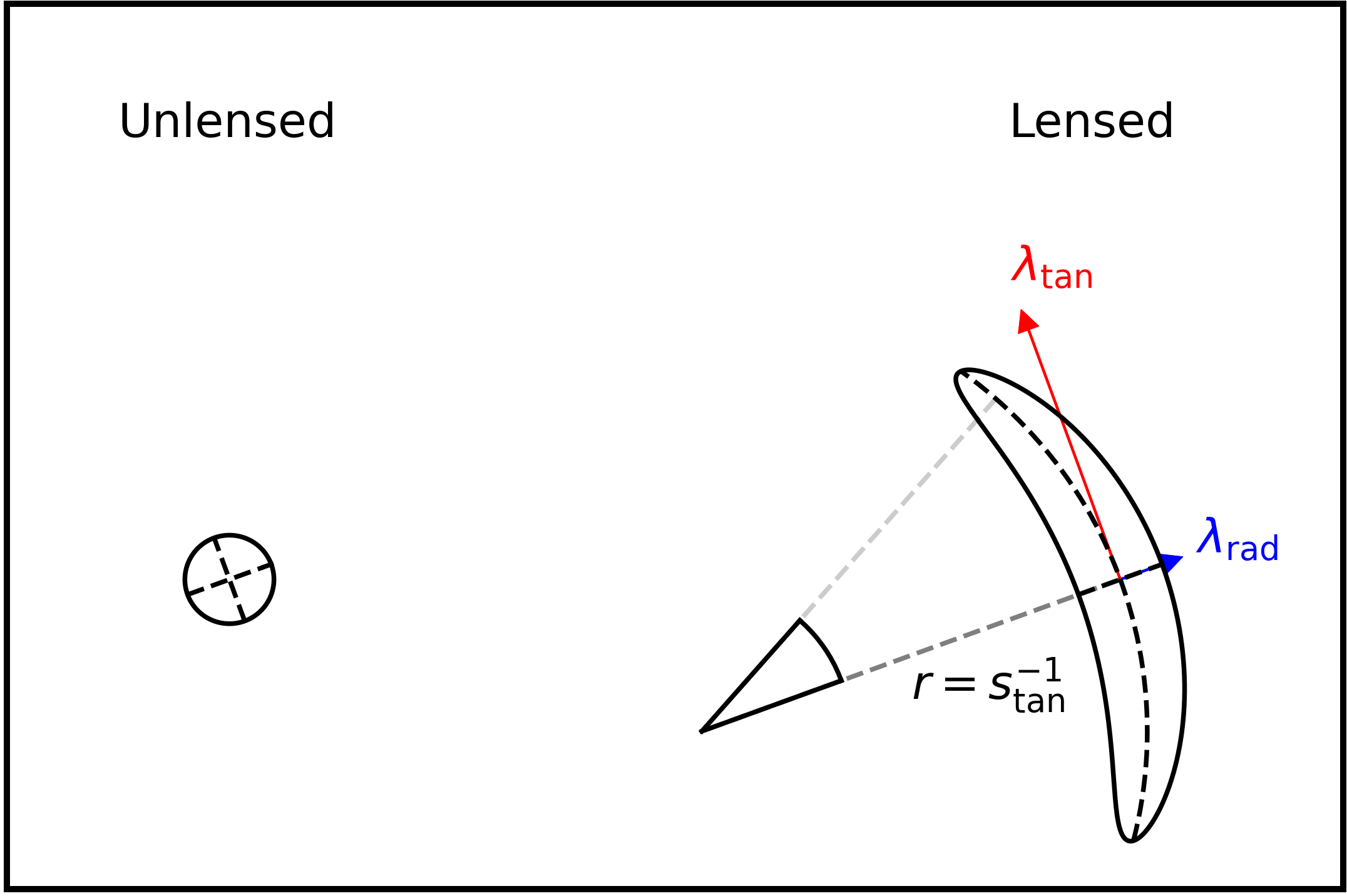}
    \caption{Unlensed image of a circle (on the left) and the image of the circle after it is lensed in the  curved arc basis with parameters $\lambda_\mr{tan} = 6$, $\lambda_\mr{rad} = 1$, $s_\mr{tan} = 0.1$ (on the right).}
    \label{fig:curved_arc_illus}
\end{figure}

%%%%%%%%%%%%%%%%%%%%%%%%%%%%%%%%%%%%%%%%%%%%%%%
\subsection*{Angular Deflections in a Curved Arc Basis}

Our lens model for the smooth angular deflections around the images can be characterized using the curved arc basis, proposed by \cite{curved_arc_theory}. Curved arc basis is a formalism to describe gravitational lensing distortion effects using the eigenvectors and eigenvalues (and their directional derivatives) of the local lensing Jacobian. The image of a circular object gets lensed into a curved arc, as shown in Fig. \ref{fig:curved_arc_illus}. This lens mapping is described by a set of parameters: $s_\mr{tan}$, the inverse of the curvature radius; $\lambda_\mr{tan}$, the tangential stretch; $\lambda_\mr{rad}$, the radial stretch; and $\phi$, the orientation defined as the angle between the tangential stretching axis and the $x$ axis. The angular deflection field of a curved arc basis is equivalent to that of a singular isothermal sphere (SIS) and a mass sheet transformation (MST). More explicitly, the angular deflection can be written as
\begin{equation}\label{eq:curved_arc_equation}
    \vec \alpha(\vec \theta) = \lambda^{-1}_\mr{rad} \left[\vec \alpha_\mr{SIS}(\vec \theta) - \vec \alpha_\mr{SIS}(\vec \theta_0)\right] + (1-\lambda^{-1}_\mr{rad})(\vec \theta - \vec \theta_0).
\end{equation}
The SIS angular deflection, meanwhile, is given by
\begin{equation}
    \vec \alpha_\mr{SIS} = s_\mr{tan}^{-1} \left(1- \frac{\lambda_\mr{rad}}{\lambda_\mr{tan}}\right)\left(\frac{\vec \theta - \vec \theta_c}{|\vec \theta - \vec \theta_c|}\right),
\end{equation}
where $\vec \theta_c$ is the centroid of the curvature radius. For derivations of these equations and detailed descriptions of curved arc basis properties, see \cite{curved_arc_theory}. The curved arc basis is formulated this way to make sure that the location $\vec \theta_0$ gets mapped to itself with $\vec \alpha(\vec \theta_0) = 0$. Along the curvature radius, the tangential and the radial stretch due to a curved arc are constant. In the limit $s_\mr{tan} \rightarrow \infty$, a curved arc simplifies to a constant magnification and shear across the image. \cite{linear_cluster_lensing} has demonstrated that, in certain cases when the extent of the lensed images is small, one can model multiple images in clusters by using this constant shear and magnification limit. Curved arc basis has a broader applicability, especially for highly magnified and extended arcs, by having a freely varying curvature that is not fully captured by fitting to an ellipse with specified major and minor axes.

An important degeneracy emerges between the source size and the lens model magnification when the lensing field is modeled locally. If one scales the source by a constant factor $a$ and simultaneously scales the tangential and radial stretches by $a^{-1}$, the image remains unchanged. To avoid this degeneracy, during our modeling, we fix $\lambda_\mr{rad} = 1$ for the lens model of the first image. The remaining smooth model parameters that we fit represent relative distortions with respect to the first image.

%%%%%%%%%%%%%%%%%%%%%%%%%%%%%%%%%%%%%%%%%%%%%%%
\subsection*{The Offset Between Images}

We center each image around its brightest pixel since the brightest pixels get lensed to approximately the same position on the source plane. The smooth model for the local angular deflection field of each image (shown in Fig. \ref{fig:real_data_fit}) consists of a curved arc basis given by Eq. \eqref{eq:curved_arc_equation} and a constant shift $\vec \alpha_i$. This shift is set to $\vec \alpha_1 = 0$ for the first image since the model only needs to make relative corrections of constant shifts between the images.

%%%%%%%%%%%%%%%%%%%%%%%%%%%%%%%%%%%%%%%%%%%%%%%
\subsection*{The Source Model}

To capture the irregular light distribution of source galaxies, we model the source light with shapelets \citep{shapelets,lenstronomy_shape,Birrer:2018vtm}. Shapelets are an orthonormal set of weighted-Hermite polynomials. Once the lens model is fixed, the pixels in the image are simply linear combinations of the coefficients of these polynomials. One can then obtain these coefficients by a simple matrix inversion, which subsequently gives the source model. The complexity of the source can be tuned by varying the shapelet-order parameter $n_\mr{max}$. The number of degrees of freedom $N$ in the source light distribution as a function of the order parameter is given by $N = (n_\mr{max}+1)(n_\mr{max}+2)/2$. The angular scale of the entire shapelet set is determined by the size-scaling parameter $\delta$. To capture both the sharp and bright features of the center of the source galaxy, as well as its extended light distribution, we model it with two sets of shapelets: one with a smaller scale parameter
$\delta_2 \in [0,0.2$"$]$, the other with a larger one $\delta_1 \in [0.2$"$,0.7$"$]$, where the intervals show the uniform priors within which they are allowed to vary. The centroids of these sets are varied independently. Unlike the shapelet coefficients mentioned earlier, which are determined by a matrix inversion, the width and the centroid of the shapelet basis are non-linear model parameters whose posteriors are obtained with nested sampling. 

%%%%%%%%%%%%%%%%%%%%%%%%%%%%%%%%%%%%%%%%%%%%%%%
\subsection*{Bayesian Inference with Nested Sampling}

If one assumes the errors of the pixel values to be Gaussian, the probability of the data being a random realization of the model reconstruction $\mathbf{M(\mathbf{q}})$ is given by
\begin{equation}
    P(\mathbf{D}|\mathbf{q}) = \frac{\exp\left[-\frac{1}{2}\left(\mathbf{D} - \mathbf{M(\mathbf{q})}\right)^T \boldsymbol{\Sigma}^{-1}_\mr{pixel} \left(\mathbf{D} - \mathbf{M(\mathbf{q})}\right) \right]}{\sqrt{(2\pi)^{\mr{dim}(\mathbf{D})} \det (\boldsymbol{\Sigma}_\mr{pixel})}},
\end{equation}
where $\boldsymbol{\Sigma}_\mr{pixel}$ is the covariance matrix of the pixel errors. It is common to assume, as we have also done in our analysis, that the pixel noise is uncorrelated.

We are interested in calculating the probability distribution of the model parameter values given the data. One can obtain the model parameter posteriors by evaluating:
\begin{equation}\label{eq:postint}
    P(\mathbf{q}|\mathbf{D}) = \int d\mathbf{q}\, P(\mathbf{q}) P(\mathbf{D}|\mathbf{q}),
\end{equation}
where $P(\mathbf{q})$ encodes the model parameters priors. To numerically evaluate the integral in Eq. \eqref{eq:postint}, we use \texttt{dynesty} \citep{2020MNRAS.493.3132S}, a publicly available nested sampling algorithm package.

%%%%%%%%%%%%%%%%%%%%%%%%%%%%%%%%%%%%%%%%%%%%%
\section{Modeling Real Data: the cluster lens SMACS 0723}
\label{sec:smacs}

In this section, we discuss our analysis of the three images of the same background galaxy, lensed by the foreground galaxy cluster SMACS J0723.3-7327 (hereafter SMACS 0723, see Fig. \ref{fig:smac_wide}). We have chosen to study this cluster lens because it provides large, bright, and extended multiply-imaged sources in addition to having deep JWST imaging providing high sensitivity. We have chosen this particular source galaxy to model because it is one of the most luminous lensed arcs in this lensing system, and its multiple images are in locations with little contamination from other objects. This cluster was first discovered by \cite{smacs_discovery}, and has since then been studied extensively with strong lens models using older HST data \citep{smacs_hst_model1}, as well as the more recent JWST data \citep{smacs_jwst_model1,smacs_redshifts,Pascale_2022}. These lens models were obtained by using the standard parametric cluster lensing methods mentioned earlier in \cref{sec:intro}. A redshift survey of this cluster lens reveals that member galaxies of the cluster SMACS J0723 have redshifts ranging between 0.367 - 0.408.
The source galaxy that we study here in detail has a redshift of $z=1.449$. These are published spectroscopic redshifts from \citep{smacs_redshifts}. In order to study the local lensing deflection field around the multiple images of the source galaxy, we extract three cutouts as 120$\times$120 pixel square arrays, as shown in Fig. \ref{fig:smac_wide}. 

The image (known as ``Webb's First Deep Field\footnote{\url{https://www.stsci.edu/jwst/science-execution/program-information.html?id=2736}}") is captured by the Near Infrared Camera (NIRCam) instrument on the JWST with approximately $17$ hours of exposure time. The image has a $10\sigma$ point source depth of $4.65$ nJy, which corresponds to $29.8$ in AB magnitude. It is taken with the F200W filter, which admits light of wavelength roughly between $1.75-2.25\,\mu$m. The pixel size of the reduced images is $0.031''$. The full width at half maximum (FWHM) of the point spread function (PSF) of JWST at a wavelength of $2\,\mu$m is $0.064''$. All reduced images of SMACS J0723 used in this study are publicly available on the Mikulski Archive for Space Telescopes (MAST)\footnote{\url{mast.stsci.edu/portal/Mashup/Clients/Mast/Portal.html}}.

To model the point spread function (PSF) of JWST, we use stars in the wider field of the image. Stars are effectively point sources, which makes them useful for modeling the point spread function. We use the publicly available PSF reconstruction package \textsc{psfr}\footnote{\url{https://github.com/sibirrer/psfr}} (Birrer et al. in prep).

%%%%%%%%%%%%%%%%%%%%%%%%%%%%%%%%%%%%%%%%%%%%%
\subsection*{Results of modeling the extended arc in SMACS 0723}

\begin{figure}
    \centering
    \includegraphics[width=\linewidth]{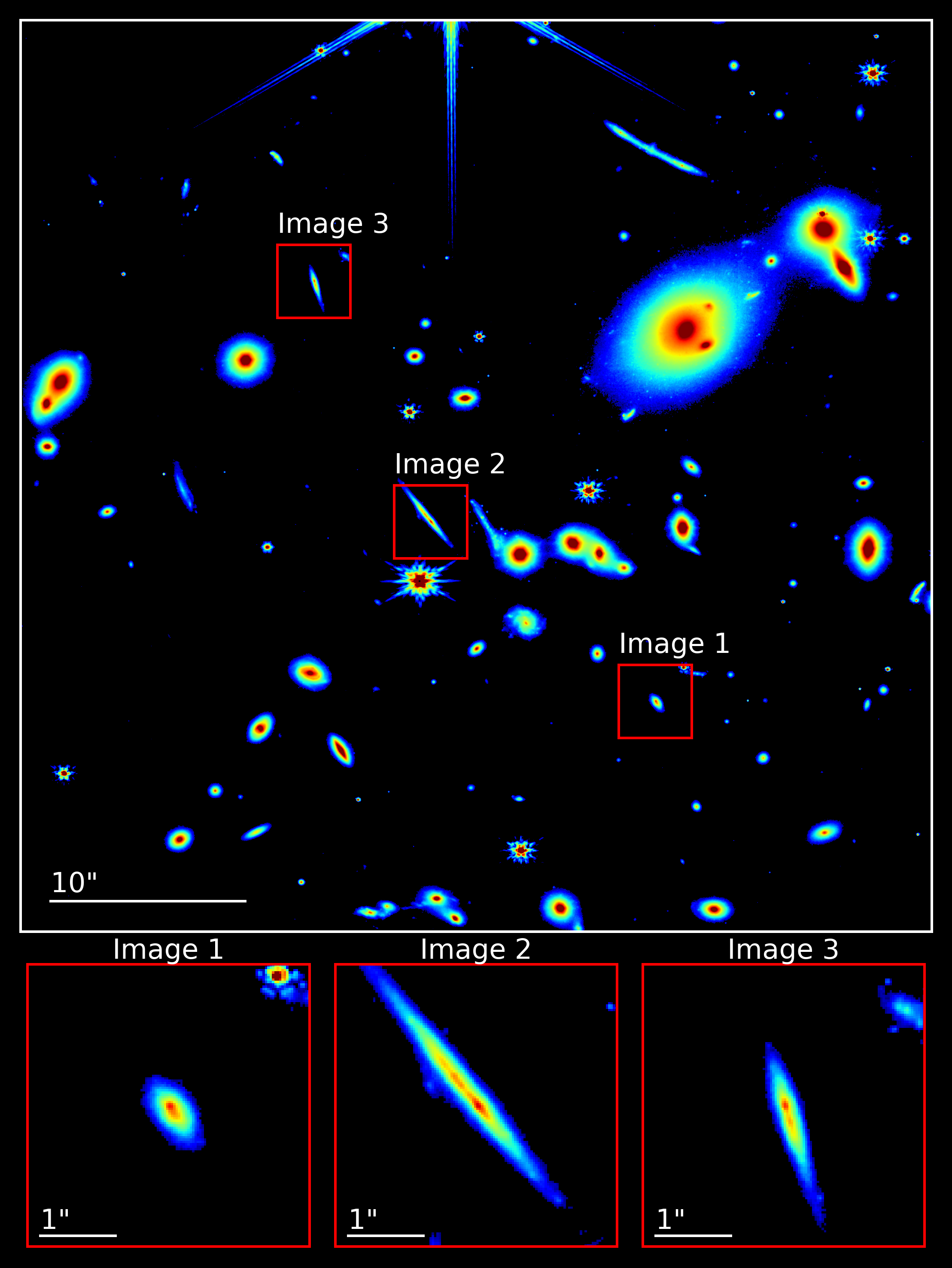}
    \caption{\textit{Top:} The wider field view of the galaxy cluster SMACS 0723. The three different images of our source of interest are shown with red squares. \textit{Bottom:} Cutouts of the three lensed images of the background source galaxy.}
    \label{fig:smac_wide}
\end{figure}

\begin{figure}
    \centering
    \includegraphics[width=\linewidth]{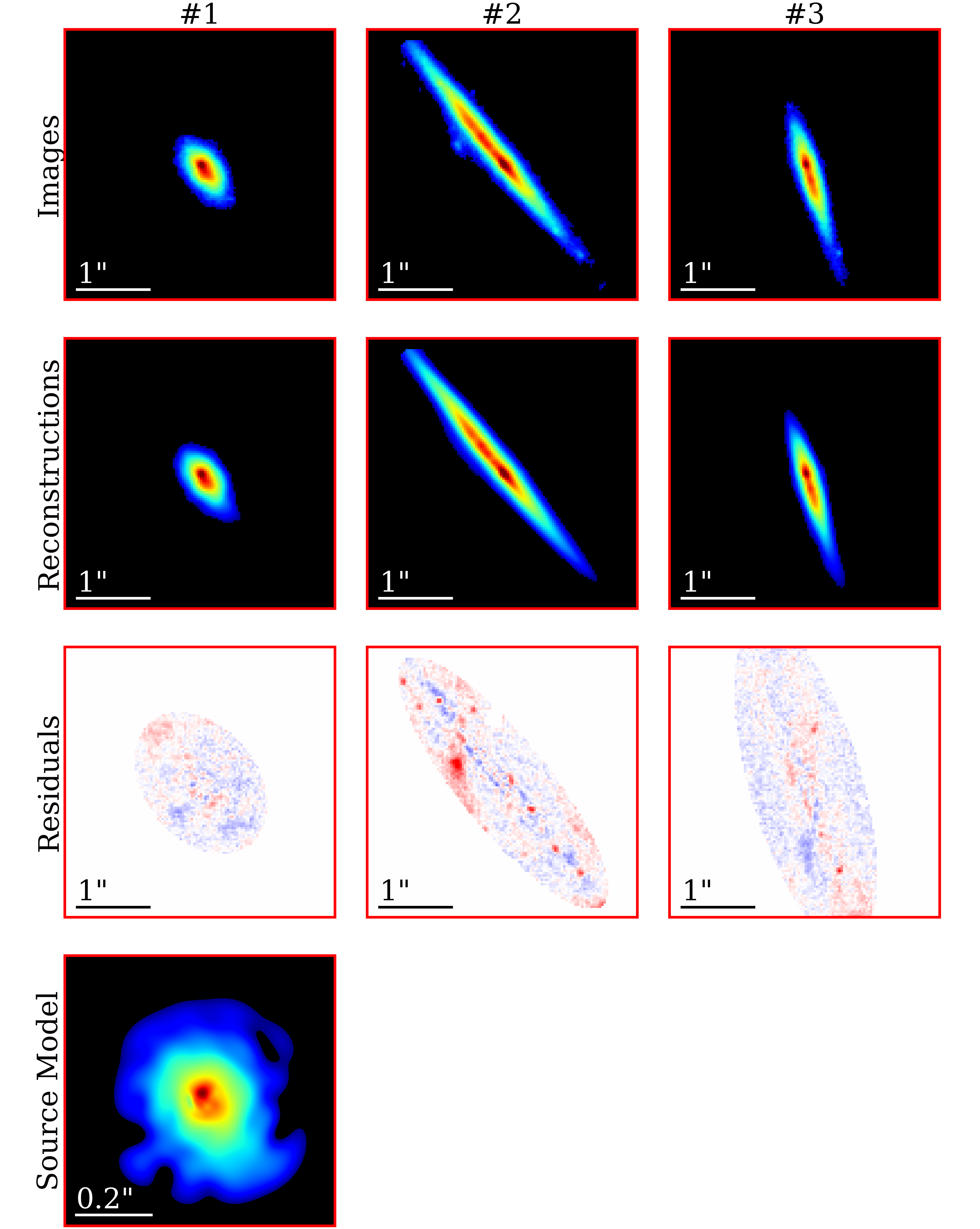}
    \caption{\textit{First row:} Background subtracted and masked images. \textit{Second row:} The reconstructions of each image by the best fit of our source and lens model. \textit{Third row:} The normalized residuals between the data and the reconstruction. \textit{Fourth row:} The source reconstruction made with shapelets.}
    \label{fig:real_data_fit}
\end{figure}

\begin{figure*}
    \centering\includegraphics[width=\linewidth]{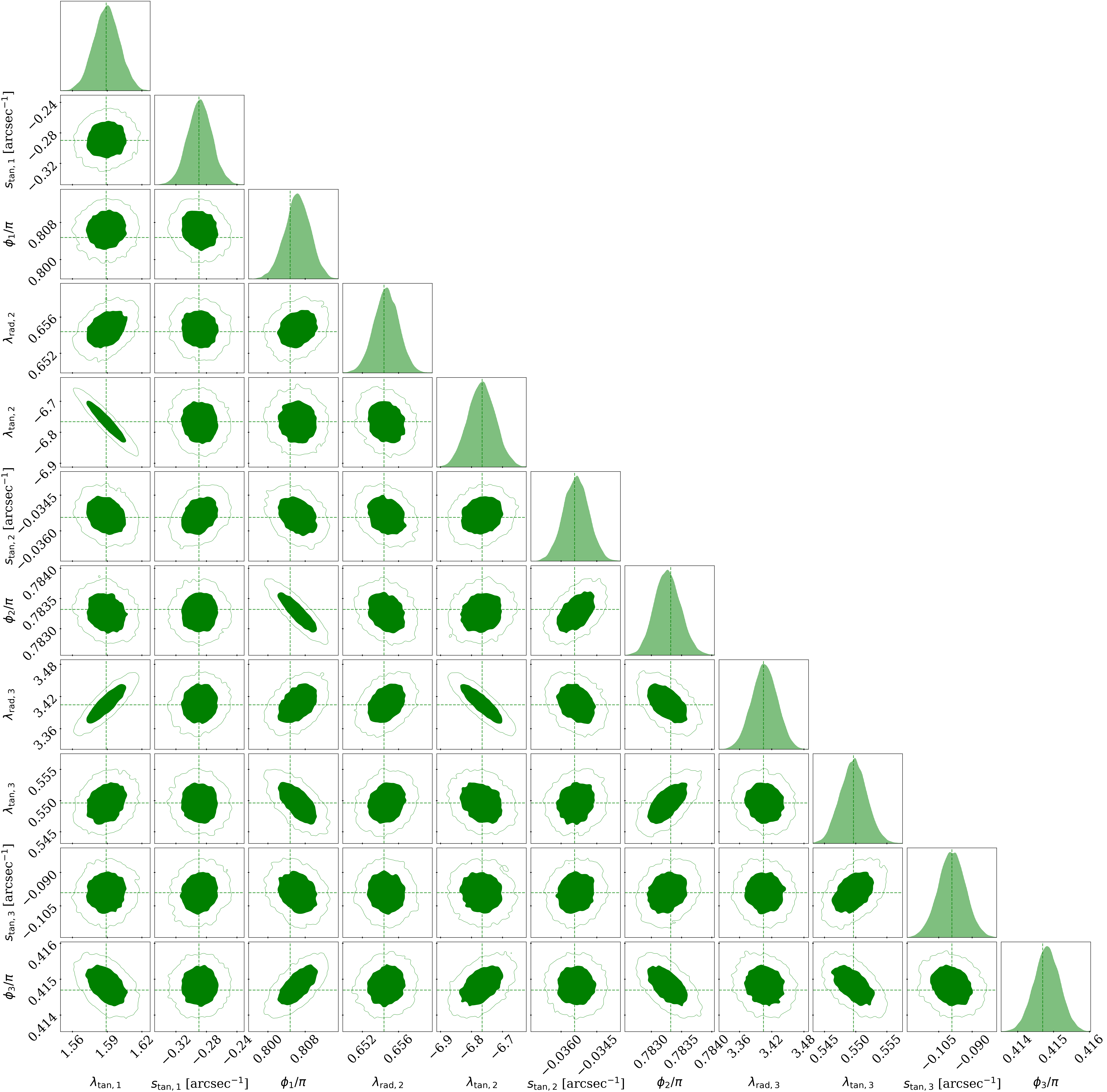}
    \caption{Posterior probability distributions of the lens model parameters listed in Table \ref{tab:table_1}. The best fits are shown as dashed lines. Parameters determining the relative offsets between the images and the source model parameters are omitted for brevity. They are uncorrelated with the parameters shown here.}  
    \label{fig:smooth_post}
\end{figure*}

We analyze the three images shown in Fig. \ref{fig:smac_wide} with a smooth lens model using the curved arc basis and a constant shift for each image. For the source model, the shapelet parameter is set to be $n_\mr{max} =8$, as this gives the lowest Bayesian Information Criterion (BIC). BIC is defined as BIC $\equiv k \ln(n) + \chi^2$, where $k$ is the number of model parameters and $n$ is the number of data points. We show in Table \ref{tab:table_1} the best fit values of the model parameters along with their $1\sigma$ uncertainties as well as the uniform prior intervals. The $1\sigma$ and $2\sigma$ contours of the posterior probability distributions of the lens model parameters are shown in Fig. \ref{fig:smooth_post}. The images, their best fit reconstructions, and residuals are shown in Fig. \ref{fig:real_data_fit}, along with the source reconstruction.

\begin{table*}
    \centering
    \begin{tabular}{||c|c|c|c||}
    \hline
    Parameter Name [units] & Measurement & Priors & Description\\
    \hline
    $\lambda_{\mathrm{tan},1}$ & $1.59 \pm 0.02$ & $[1.0,2.5]$ & Tangential stretch of image 1 \\ 
    \hline
    $s_{\mathrm{tan},1}$[arcsec$^{-1}$] & $-0.29 \pm 0.03$ & $[-1.0,1.0]$ & Arc curvature of image 1 \\ 
    \hline
    $\phi_{1}/\pi$  & $0.81 \pm 0.012$ & $[0.0,1.0]$ & Arc orientation of image 1\\ 
    \hline
    $\lambda_{\mathrm{rad},2}$ & $0.654 \pm 0.003$ & $[0.,2.0]$ & Radial stretch of image 2 \\ 
    \hline
    $\lambda_{\mathrm{tan},2}$ & $-6.77 \pm 0.09$ & $[-15.0,-3.0]$ & Tangential stretch of image 2  \\ 
    \hline
    $s_{\mathrm{tan},2}$ [arcsec$^{-1}$] & $-0.035 \pm 0.001$ & $[-1.0,1.0]$  & Arc curvature of image 2 \\ 
    \hline
    $\phi_{2}/\pi$ & $0.7832 \pm 0.0010$ & $[0.0,1.0]$  & Arc orientation of image 2\\ 
    \hline
    $\alpha_{2,x}$ [arcsec] & $0.040 \pm 0.002$ & $[-0.1,0.1]$ & x-offset of image 2\\ 
    \hline
    $\alpha_{2,y}$ [arcsec] & $-0.013 \pm 0.001$ & $[-0.1,0.1]$ & y-offset of image 2\\ 
    \hline
    $\lambda_{\mathrm{rad},3}$ & $3.41 \pm 0.05$ & $[1.0,7.0]$ & Radial stretch of image 3 \\ 
    \hline
    $\lambda_{\mathrm{tan},3}$ & $0.550 \pm 0.004$ & $[0.0,2.0]$ & Tangential stretch of image 3 \\ 
    \hline
    $s_{\mathrm{tan},3}$ [arcsec$^{-1}$] & $-0.1 \pm 0.01$ & $[-1.0,1.0]$ & Arc curvature of image 3\\ 
    \hline
    $\phi_{3}/\pi$ & $0.415 \pm 0.002$ & $[0.0,1.0]$ & Arc orientation of image 3\\ 
    \hline
    $\alpha_{3,x}$ [arcsec] & $0.030 \pm 0.001$ & $[-0.1,0.1]$ & x-offset of image 3\\ 
    \hline
    $\alpha_{3,y}$ [arcsec] & $0.001 \pm 0.001$ & $[-0.1,0.1]$ & y-offset of image 3 \\ 
    \hline
    $\delta_1$ [arcsec] & $0.42 \pm 0.01$ & $[0.2,0.7]$ & Shapelet scale parameter\\ 
    \hline
    $x_1$ [arcsec] & $0.040 \pm 0.003$ & $[-0.2,0.2]$ & Shapelet center x\\ 
    \hline
    $y_1$ [arcsec] & $0.065 \pm 0.003$ & $[-0.2,0.2]$ & Shapelet center y\\ 
    \hline
    $\delta_2$ [arcsec] & $0.061 \pm 0.001$ & $[0.0,0.2]$ & Shapelet scale parameter\\ 
    \hline
    $x_2$ [arcsec] & $0.0185 \pm 0.0015$ & $[-0.2,0.2]$ & Shapelet center x\\ 
    \hline
    $y_2$ [arcsec] & $0.0465 \pm 0.0015$ & $[-0.2,0.2]$ & Shapelet center y\\ 
    \hline
    \end{tabular}
    \caption{Mean and $1\sigma$ uncertainties of the lens-model parameters used to analyze three images of the same source shown in Fig. \ref{fig:real_data_fit}. See \cref{sec:smacs} for a detailed description of the lens model.}
    \label{tab:table_1}
\end{table*}

%%%%%%%%%%%%%%%%%%%%%%%%%%%%%%%%%%%%%%%%%%%%%
\section{Mock Tests with JWST quality data}
\label{sec:JWSTsims}

In this section, we describe several tests that we ran on mock data. Data properties such as PSF, exposure, noise, and resolution are all set to be identical to the JWST image of the cluster lens SMACS 0723 analyzed earlier, to make the mock data sets as realistic as possible. Our goal here is to quantify the sensitivity that JWST-type observations of lensed arcs in clusters give us in measuring individual low-mass perturber properties.

%%%%%%%%%%%%%%%%%%%%%%%%%%%%%%%%%%%%%%%%%%%%%
\subsection{Probing Mock Perturber Properties}

\begin{figure}
    \centering
    \includegraphics[width=\linewidth]{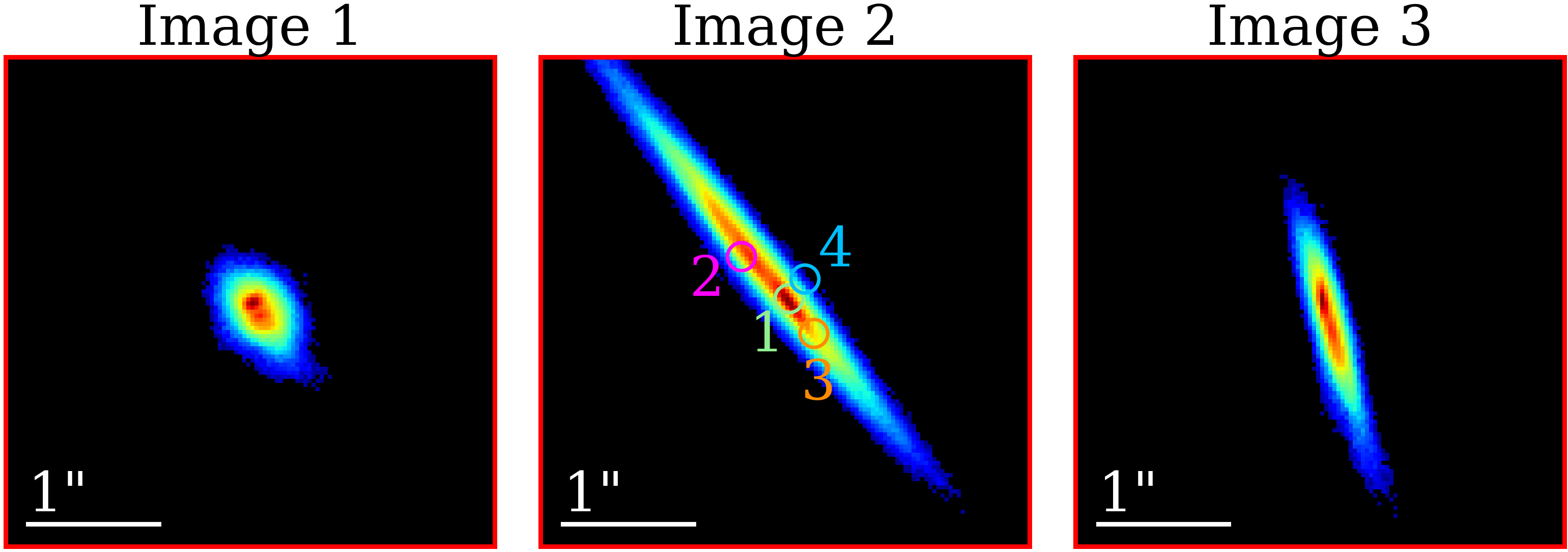}
    \caption{An example of one of the mock data sets that are used in \cref{sec:JWSTsims}. The numbered and colored circles show the angular positions of the perturbers that are placed in four different data sets, where each data set only has one perturber. The locations are in decreasing order of surface brightnesses of 75\%, 50\%, 25\%, and 5\% of the brightest pixel.}  
    \label{fig:mock}
\end{figure}

We use the best fits of the smooth-lens model parameters that we obtained from our analysis of SMACS 0723, as well as the source reconstruction, to create multiple mock data sets.  Each of the data sets consists of three images of the same source, unless otherwise specified. One such data set is shown in Fig. \ref{fig:mock}. In each data set, we have placed a perturber near one of the bright images of the background source. The parameters of these perturbers are set to be different for each data set in order to investigate our ability to probe different properties. We then obtain the posterior probability distributions of these individual perturber parameters with a nested sampling algorithm. The lens model that we use to analyze these data sets is an extension of the smooth model described in \cref{sec:smacs}. In addition to the curved arc bases that capture the smooth component of the angular deflection field locally, we include an individual perturber whose mass, position, and various other properties are allowed to vary freely. The perturber is set to be at the same redshift as the cluster, unless otherwise specified.

\begin{figure}
    \centering
    \includegraphics[width=\linewidth]{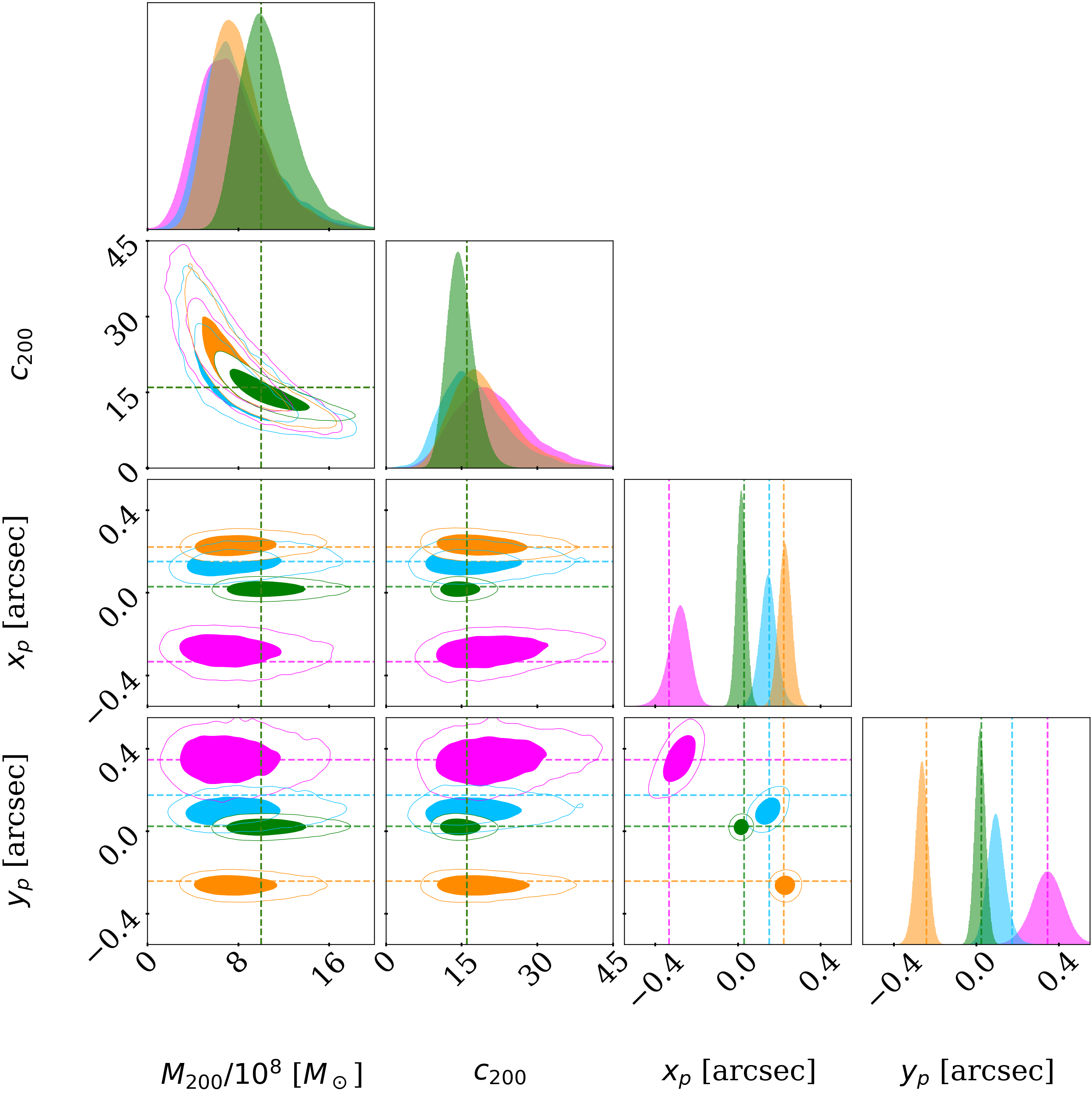}
    \caption{Posterior probability distributions of the perturber parameters for the four different data sets with different perturber positions. The colors match the colors used in Fig. \ref{fig:mock} for each perturber.} 
    \label{fig:mock_post1}
\end{figure}

%%%%%%%%%%%%%%%%%%%%%%%%%%%%%%%%%%%%%%%%%%%%%
\begin{figure}
    \centering
    \includegraphics[width=\linewidth]{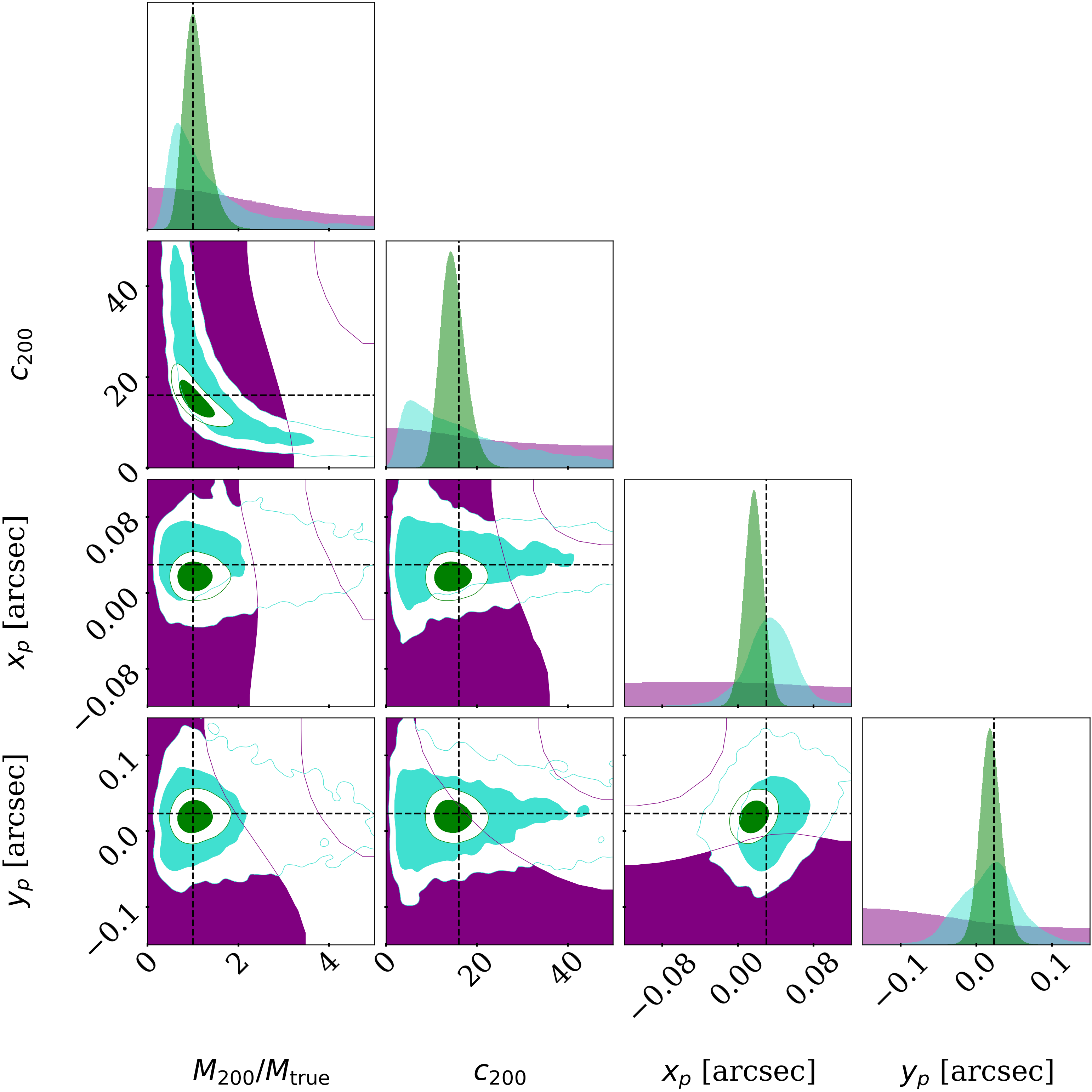}
    \caption{Posterior probability distributions of the NFW perturber parameters for the three different data sets with different masses. The \textcolor{ForestGreen}{green} contours correspond to ``perturber 1" with mass $10^{9}\,\msun$, which is also shown in Fig. \ref{fig:mock_post1}. The \textcolor{Turquoise}{turquoise} contours correspond to a perturber with mass $10^{8}\,\msun$, and the \textcolor{violet}{purple} contours correspond to a perturber with mass $10^{7}\,\msun$, both placed at the same location as ``perturber 1".}  
    \label{fig:mock_post_lowmass}
\end{figure}

\begin{figure}
    \centering
    \includegraphics[width=\linewidth]{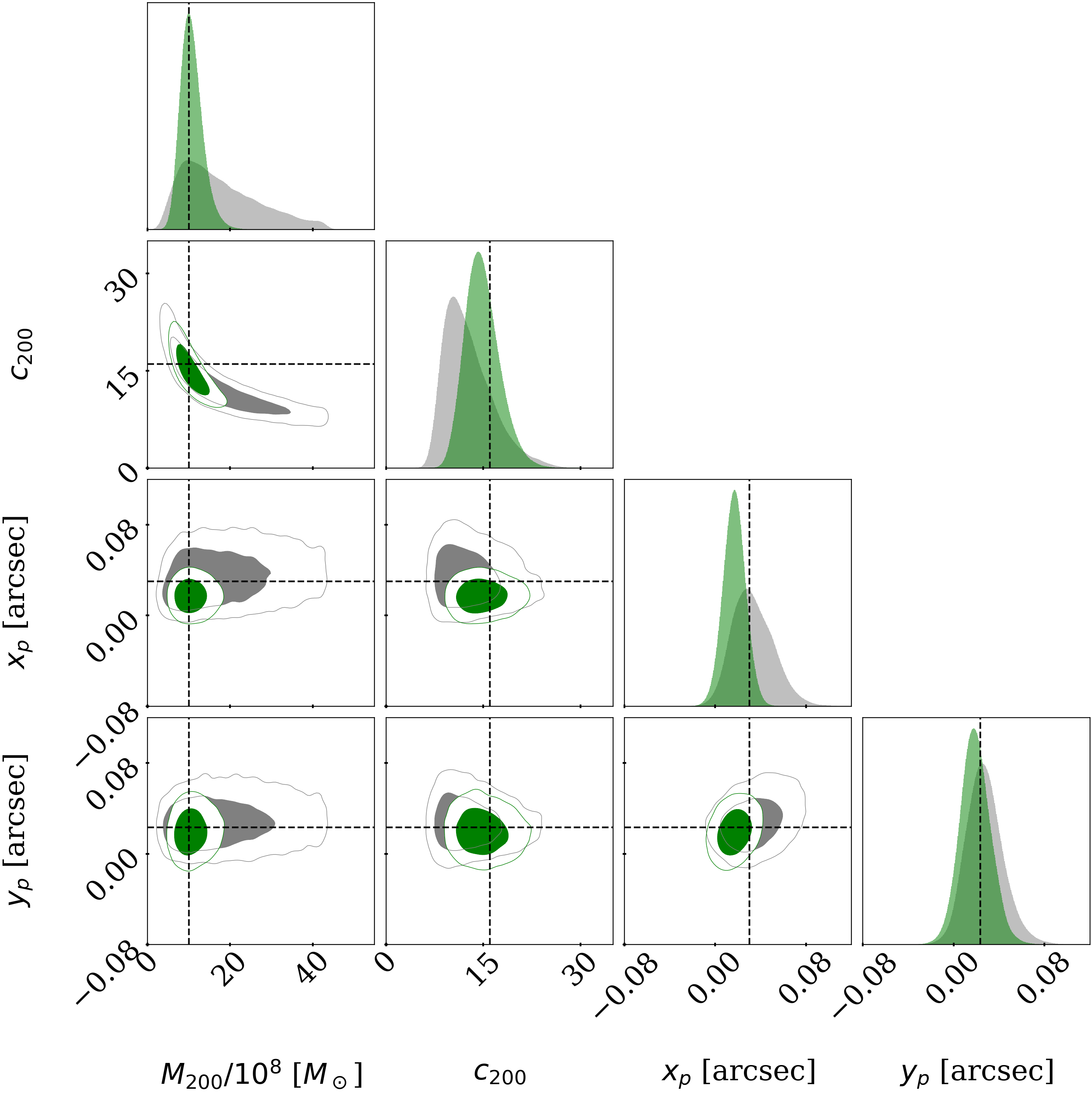}
    \caption{Posterior probability distributions of the perturber parameters for the two different data sets with different arc magnifications. The \textcolor{ForestGreen}{green} contours correspond to ``perturber 1", which is also shown in Figs. \ref{fig:mock_post1} and \ref{fig:mock_post_lowmass}. The \textcolor{gray}{gray} contours correspond to a perturber with the same properties as ``perturber 1" but placed on top of an arc whose magnification is 50\% less than that of ``perturber 1".}  
    \label{fig:mock_post_lowmag}
\end{figure}

%%%%%%%%%%%%%%%%%%%%%%%%%%%%%%%%%%%%%%%%%%%%%
\subsection{Detecting Individual Perturbers}

%%%%%%%%%%%%%%%%%%%%%%%%%%%%%%%%%%%%%%%%%%%%%
\subsubsection{Perturber Position}

The first property that we investigate is our ability to detect a perturber based on its angular proximity to the bright regions in the images. In general, we expect our ability to measure perturbers to increase the closer they are to sharp and bright features in the lensed images. This is because the changes in pixel brightness values in an image due to the very small angular deflections of a low-mass perturber are proportional to the intensity and gradient of the surface brightness of the source. The perturbers are assumed to have a Navarro-Frenk-White (NFW) \citep{NFW,NFW2} density profile given by
\begin{equation}\label{eq:nfw}
    \rho_\mr{NFW}(r) = \frac{\rho_\mr{s}}{\dfrac{r}{r_\mr{s}}\left(1 + \dfrac{r}{r_\mr{s}}\right)^2},
\end{equation}
where $r_\mr{s}$ is the scale radius and $\rho_\mr{s}$ is the density at the scale radius. We parametrize the NFW halo by its mass $M_{200}$ and concentration $c_{200}$. The former is defined as the mass within the radius $r_{200}$ where the average density of the halo is 200 times the critical density of the Universe, and the latter is defined as $r_{200}/r_\mr{s}$.

For this test, we place a single perturber in four different data sets, and we choose the location of the perturber randomly, with the condition that the surface brightness at the location needs to be 75\%, 50\%, 25\%, and 5\% of the brightest pixel in the image. The perturber positions explored are shown in Fig. \ref{fig:mock}, labeled with different numbers and colors. We find that we are able to detect the perturber and recover its parameters for all four cases. We show the posterior probability distributions of the model parameters in Fig. \ref{fig:mock_post1}. We find that the lowest errors are for ``perturber 1", which is placed closest to the bright and sharp central region of the image, giving us the highest constraining power. For the other perturber locations vis-à-vis the brightest pixel in the image, we see that our sensitivity is not a simple function of the surface brightness of the image.

%%%%%%%%%%%%%%%%%%%%%%%%%%%%%%%%%%%%%%%%%%%%%
\subsubsection{Lowest Detectable Perturber Mass}

In order to quantify the lowest mass that we can detect with this particular extended arc configuration, we test two other cases with the same optimal position as ``perturber 1" above, but with masses $10^{8}\,\msun$ and $10^{7}\,\msun$ respectively. We show the posterior probability distributions of the model parameters in Fig. \ref{fig:mock_post_lowmass}. We find that the lowest mass that we are sensitive to for optimal assumed location vis-à-vis the brightest pixel in the image is between $10^{7}\,\msun$ and $10^{8}\,\msun$. We expect this sensitivity to decrease for perturbers located at sub-optimal locations further away.

%%%%%%%%%%%%%%%%%%%%%%%%%%%%%%%%%%%%%%%%%%%%%
\subsubsection{The Effect of Source Magnification on Detection Sensitivity}\label{subsubsec:sensitiv}

The effect of a low-mass perturber is more pronounced when the magnification from the smooth component of the lensing field is already quite large. Small deviations in highly extended and magnified arcs give us better sensitivity to perturber parameters. In most cluster lenses, the multiple images are not always as highly magnified as the one we modeled in SMACS 0723. To study our ability to detect a perturber and measure its properties, we create a mock data set with the same source as before, but with 50\% lower image magnification. We do this by decreasing the tangential stretching $\lambda_\mr{tan}$ by a factor of $2$. We place a perturber at the same position as ``perturber 1", with identical mass and concentration parameter. In Fig. \ref{fig:mock_post_lowmag} we compare the posterior probability distributions of the perturber parameters for the mock data with full magnification and the lower magnification. We see that the position, concentration, and in particular the mass is much less well-constrained if the perturber is placed near an arc that is much less magnified.

%%%%%%%%%%%%%%%%%%%%%%%%%%%%%%%%%%%%%%%%%%%%%
\subsection{Measuring Perturber Properties}

For the rest of the tests in this section, the perturber is randomly placed in a location that is at least 75\% as bright as the brightest pixel in the image. We quantify our ability to recover internal properties of the perturber, computed for a perturber mass of $10^9\,\msun$.

%%%%%%%%%%%%%%%%%%%%%%%%%%%%%%%%%%%%%%%%%%%%%
\newpage
\subsubsection{Core Radius of Perturber}

\begin{figure}
    \centering
    \includegraphics[width=\linewidth]{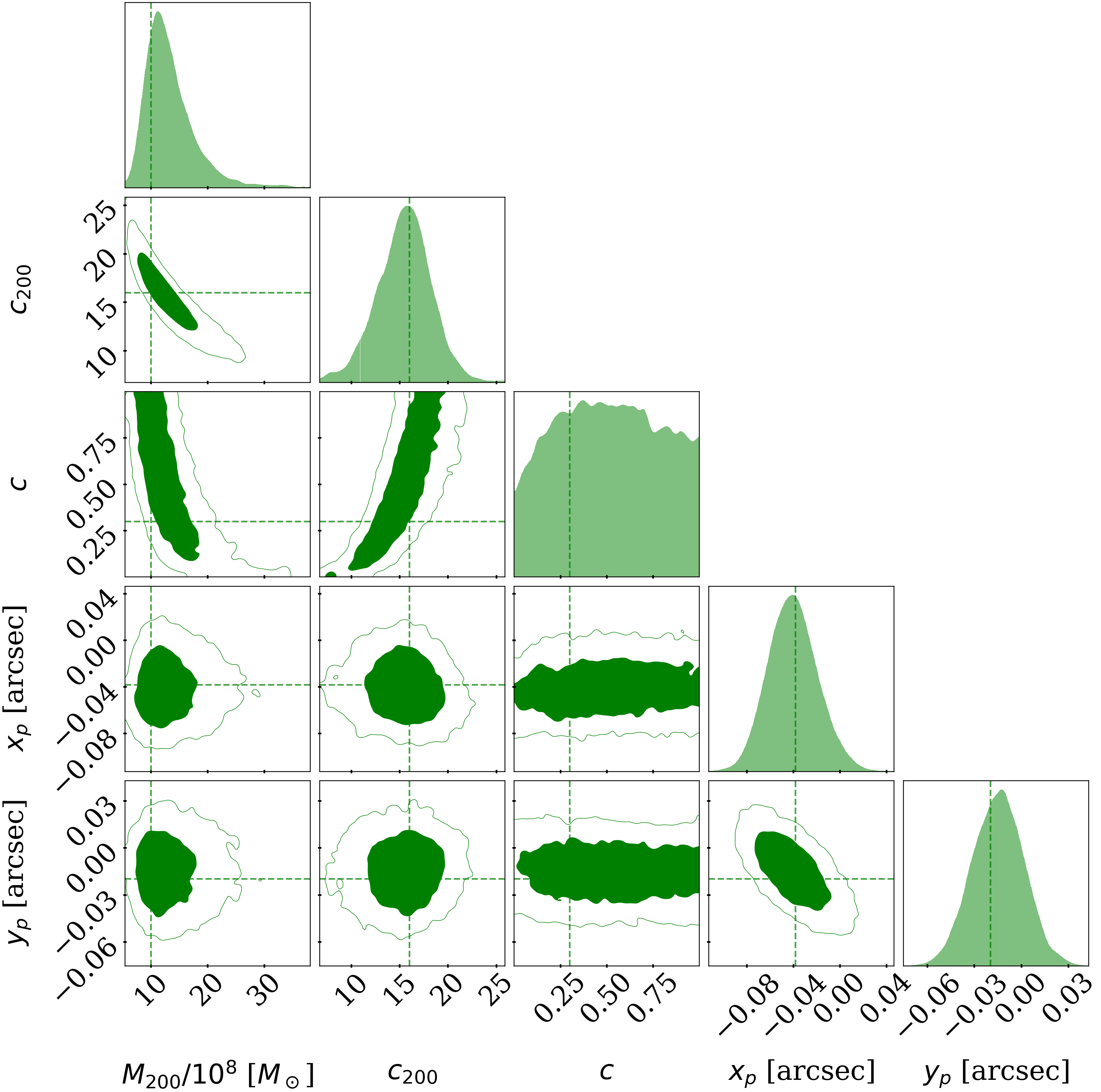}
    \caption{Posterior probability distributions of the perturber parameters for a model with a cored NFW profile. The core radius $c$ modifies the density profile as described by Eq. \eqref{eq:cnfw}.}
    \label{fig:core}
\end{figure}

We first investigate our sensitivity to measure the inner density profile shape of a perturber by using a cored NFW (cNFW) profile \citep{cnfw}, given by
\begin{equation}\label{eq:cnfw}
    \rho_\mr{cNFW}(r) = \frac{\rho_\mr{s}}{\left(\dfrac{r}{r_\mr{s}}+c\right)\left(1 + \dfrac{r}{r_\mr{s}}\right)^2},
\end{equation}
which has a core radius $r_c = c r_s$, where $c<1$. For radii smaller than  $r_c$, the density of cNFW flattens to a constant value as opposed to the divergent ``cusp" of a regular NFW profile. We create and analyze a mock data set that has a cNFW perturber with mass $M_{200} = 10^{9}\,\msun$, concentration $c_{200}=16$, and core size $c = 0.3$. In Fig. \ref{fig:core} we show the posterior probability distributions of the perturber parameters. We see that while we recover the mass, concentration, and angular position of the perturber reasonably well, we are insensitive to its core size, which is allowed to vary within $c\in [0,1]$ with a uniform prior. 

\cite{effective_power_law} have shown that with a given resolution and noise level, the observable changes in the surface brightness due to NFW subhalos can be well approximated by that of a power-law profile subhalo. We are sensitive to the amplitude and slope of this effective power-law profile within a radius range that we called region of maximum observability (RMO). 
At our noise and resolution, the angular deflections due to a cored NFW perturber with a core radius of 0.3$r_s$ and with a concentration of 16 are indistinguishable from that of an NFW perturber with a concentration of 8. This is because both of these mass distributions have nearly identical effective power law behavior in the RMO. This degeneracy between the core radius, concentration and mass can be seen in Fig. \ref{fig:core}. To probe the innermost regions of these objects we need a combination of very high resolution images, sources with sharp and bright features, and an optimal perturber position of being really close to such sharp features.

%%%%%%%%%%%%%%%%%%%%%%%%%%%%%%%%%%%%%%%%%%%%%
\subsubsection{Ellipticity of Perturber}

Cosmological N-body simulations predict dark matter halos to have significant ellipticities \citep{2002ApJ...574..538J, 2015MNRAS.449.3171B,ellipticity_sim}. We investigate our sensitivity to the ellipticity of the perturber by using an elliptical NFW (eNFW) profile \citep{elliptical_nfw}, whose angular deflection field is given by
\begin{equation}
    \vec \alpha_\mr{eNFW}(\vec x) = 
    \begin{pmatrix}
    \alpha_\mr{NFW}(x_\epsilon) \sqrt{e_1}\cos \psi_\epsilon \\
    \alpha_\mr{NFW}(x_\epsilon) \sqrt{e_2}\sin \psi_\epsilon \\
    \end{pmatrix}.
\end{equation}
Here, $e_1$ and $e_2$ are the cartesian ellipticity parameters, $\alpha_\mr{NFW}$ is the angular deflection of a spherical NFW profile, $x_\epsilon \equiv \sqrt{e_1 x_1^2 + e_2 x_2^2}$, and $\psi \equiv \arctan\left(\frac{x_1}{x_2}\sqrt{\frac{e_1}{e_2}}\right)$. The ellipticity is implemented at the level of the lensing potential by setting $\varphi_\mr{eNFW}(x) \equiv \varphi_\mr{NFW}(x_\epsilon)$. 

\begin{figure}
    \centering
    \includegraphics[width=\linewidth]{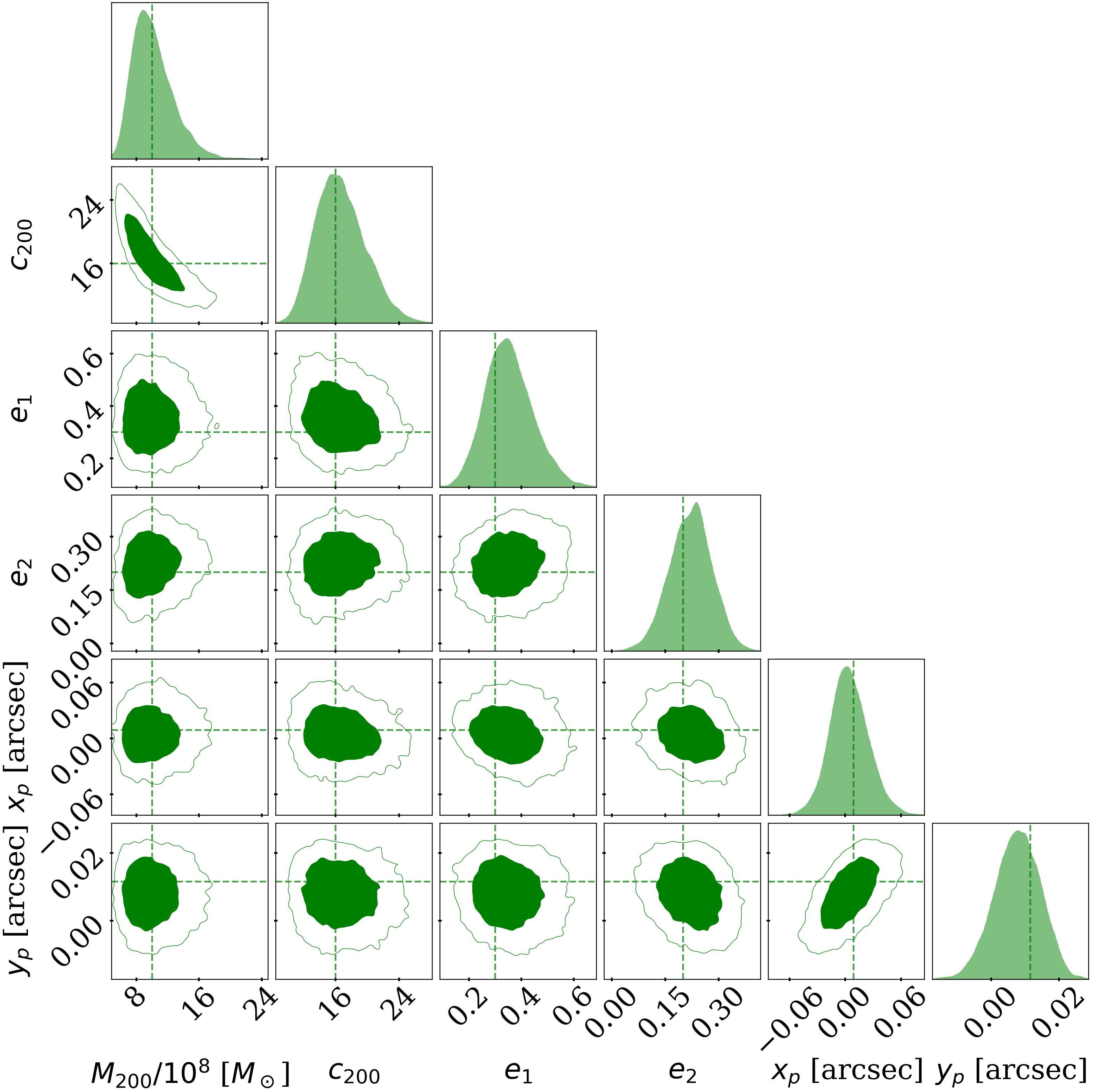}
    \caption{Posterior probability distributions of the perturber parameters for a model with a elliptical NFW profile.}
    \label{fig:ellip}
\end{figure}

We create and analyze a mock data set that has an eNFW perturber with mass $M_{200} = 10^{9}\,\msun$, concentration $c_{200}=16$, and ellipticity parameters $e_1 =0.3$; $ e_2=0.2$. We show in Fig. \ref{fig:ellip} the posterior probability distributions of these model parameters. We see that the $1\sigma$ uncertainty for $e_1$ and $e_2$ is roughly $~0.15$. The spherical case, which corresponds to $e_1 = e_2 = 0$, can be ruled out with $4.2\sigma$. The ellipticity parameters show no significant correlation with the other model parameters.

%%%%%%%%%%%%%%%%%%%%%%%%%%%%%%%%%%%%%%%%%%%%%
\subsubsection{Redshift of the Perturber}

The line-of-sight volume is expected to be full of halos that can be more numerous per unit area than the substructures within the cluster \citep{los_cluster}. Given that low-mass structures are expected to be dark per $\Lambda$CDM prediction, it is currently not possible to obtain spectroscopic measurements for a given low-mass halo to confirm its redshift. Moreover, the mass and the redshift of low-mass perturber in strong lensing are degenerate, as shown in \cite{mass_redshift_degen}. However, this degeneracy is not perfect. Specifically, the angular deflections caused by a line-of-sight halo have a non-zero curl term which vanishes for subhalos. \cite{los_detection} exploited this fact to measure the redshift of a perturber for the case of galaxy-galaxy lensing. The same principle applies for this study, as the curved arc basis is curl-free \cite{curved_arc_theory}.

\begin{figure}
    \centering
    \includegraphics[width=\linewidth]{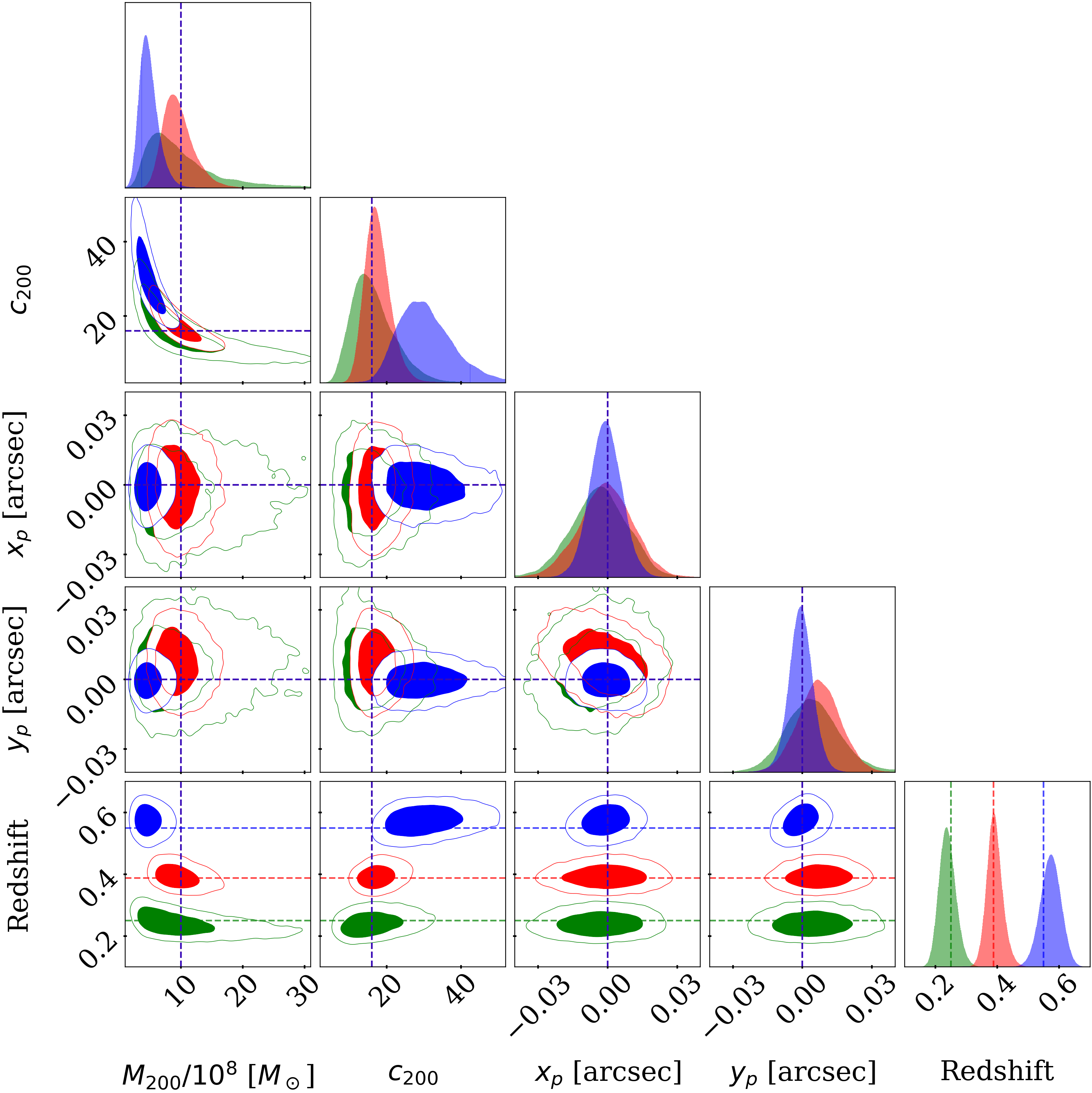}
    \caption{Posterior probability distributions of the perturber parameters for a model with a freely varying perturber redshift. The \textcolor{ForestGreen}{green} contours correspond to the foreground line-of-sight perturber with a true redshift of $0.25$. The \textcolor{red}{red} contours correspond to the subhalo perturber at the lens redshift of $0.3877$. The \textcolor{blue}{blue} contours correspond to the background line-of-sight perturber with a true redshift of $0.55$. The true values are shown as dashed lines, and the only one that varies between the three different data sets is the redshift.}
    \label{fig:redshifttest}
\end{figure}

We test how well we can measure the redshift of a perturber in a cluster lens by creating and analyzing three mock data sets, each of which has a spherical NFW perturber with mass $M_{200} = 10^{9}\,\msun$ and concentration $c_{200}=16$. The redshift of the perturber in each data set is set to be $0.25,0.3877,$ and $0.55$, respectively. The middle redshift is the average redshift of the cluster, which means that the perturber is a subhalo within the cluster. The other two are a foreground and a background line-of-sight halo. We show in Fig. \ref{fig:redshifttest} the posterior probability distributions of the model with a freely varying redshift for the perturber in each data set. We see that we are able to robustly measure the redshift of a perturber for both background and foreground line-of-sight halos, as well as for a cluster member subhalo. However, we predict a concentration that is biased higher and a mass that is lower than the true value for the background perturber at redshift $0.55$. 

%%%%%%%%%%%%%%%%%%%%%%%%%%%%%%%%%%%%%%%%%%%%%
\subsection{Detecting Wandering Black Holes}

A substantial population of wandering supermassive black holes (SMBH) is predicted to inhabit clusters \citep{wandering_bh} from high-resolution cosmological simulations in a $\Lambda$CDM cosmology. Our method offers a powerful way to probe their existence using their putative lensing signal. Here, we investigate our sensitivity to SMBHs using mock data sets. The lensing effect of a wandering SMBH, a very compact object, is expected to be quite strong, especially if the angular position of the SMBH is near one of the extended arcs \citep{wandering_bh_lensing}.  \cite{wandering_bh_lensing} note that SMBHs primarily imprint detectable signatures in rare, higher-order strong lensing image configurations but do not produce any detectable statistically significant effects in either the overall magnification profile or the integrated shear profile of the cluster. \cite{wandering_bh_lensing} report the following detectable lensing effects, where a SMBH: (i) can cause image splitting, leading to the production of additional lensed images; (ii) can introduce asymmetries in the position and magnification of multiple images; and (iii) can lead to the apparent disappearance of lensed counter-images. Of these, they predict that image splitting inside the cluster's tangential critical curve is the most prevalent and feasibly observationally detectable signature. Here, we investigate the ability to recover the presence of an individual SMBH that lies close to an extended lensed arc.

The angular deflection due to a point mass located at $\vec x=0$ is given by
\begin{equation}\label{eq:point}
    \vec \alpha(\vec x) =\frac{\theta_E^2}{x^2} \vec x.
\end{equation}
The Einstein radius $\theta_E$ is given by
\begin{equation}
    \theta_E \equiv \sqrt{\frac{4GM}{c^2} \frac{D_{ls}}{D_{l}D_{s}}},
\end{equation}
where $M$ is the mass of the deflector, $D_{l}$, $D_{s}$ and $D_{ls}$, are the angular diameter distance between the observer and the lens plane, the observer and the source plane, and the lens plane and the source plane, respectively.

\begin{figure}
    \centering
    \includegraphics[width=0.8\linewidth]{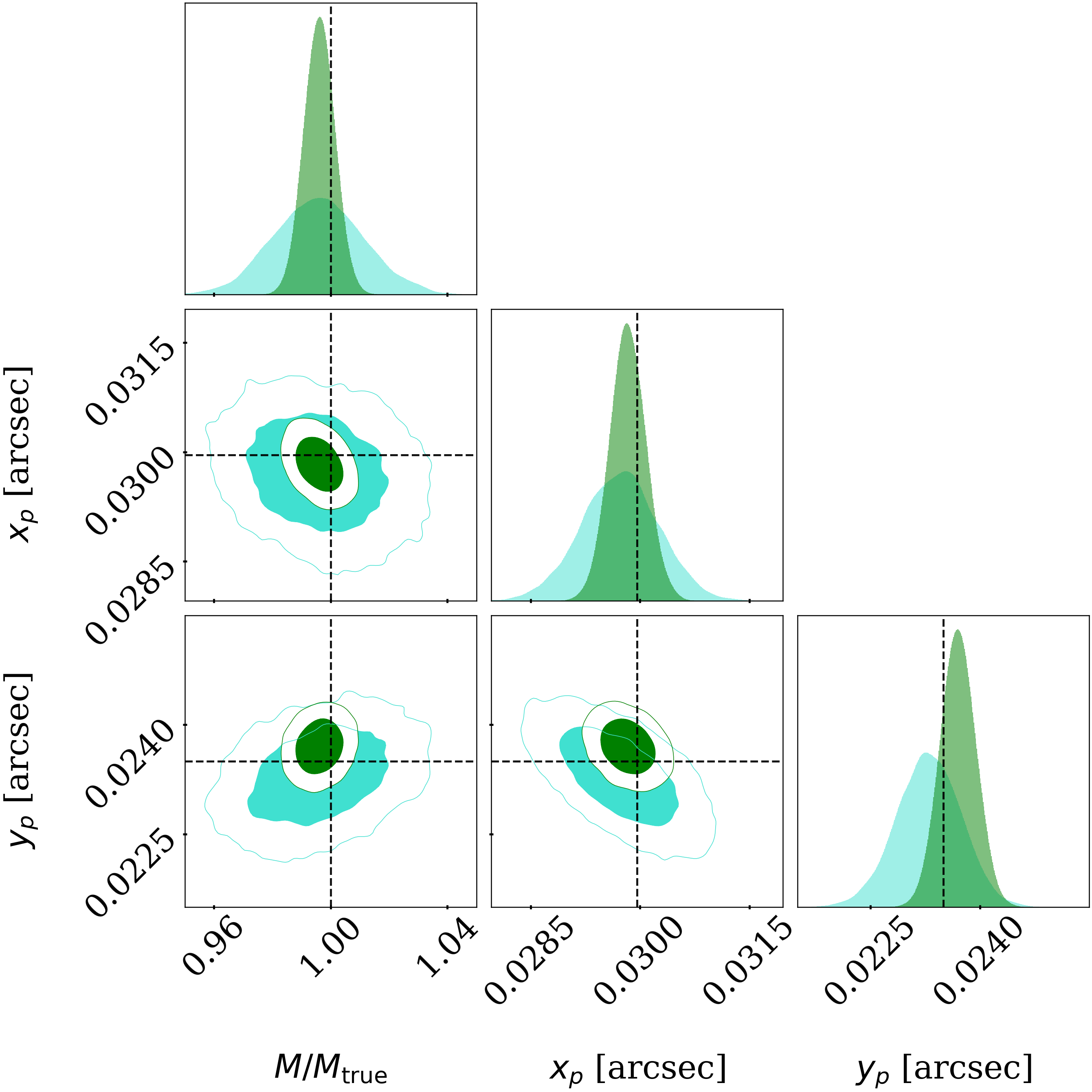}
    \caption{Posterior probability distributions of the SMBH parameters for two different data sets with different masses. The \textcolor{ForestGreen}{green} contours correspond to an SMBH with mass $10^{9}\,\msun$. The \textcolor{Turquoise}{turquoise} contours correspond to a SMBH with mass $10^{8}\,\msun$. Both are placed at the same optimal location as ``perturber 1".}
    \label{fig:bh_heavy}
\end{figure}

We create mock data sets, each with a SMBH placed at the same angular position as "perturber 1" (shown in Fig. \ref{fig:mock}) with masses of $10^9$, $10^{8}$, and $10^{7}\, \msun$. We then analyze these data sets with a point mass perturber lensing model given in Eq. \eqref{eq:point}. We show in Fig. \ref{fig:bh_heavy} the posterior probability distributions of the mass and position of the heavier black holes with masses of $10^9$ and $10^{8} \, \msun$. We see that for these heavier black holes, we are able to recover their masses with very high accuracy and precision. The uncertainties in the mass and position are roughly an order of magnitude smaller than those of the extended NFW perturbers shown in Fig. \ref{fig:mock_post_lowmass}. Because a black hole is a point mass, its lensing signal is much stronger than that of an NFW subhalo, which has an extended mass distribution.

\begin{figure}
    \centering
    \includegraphics[width=0.8\linewidth]{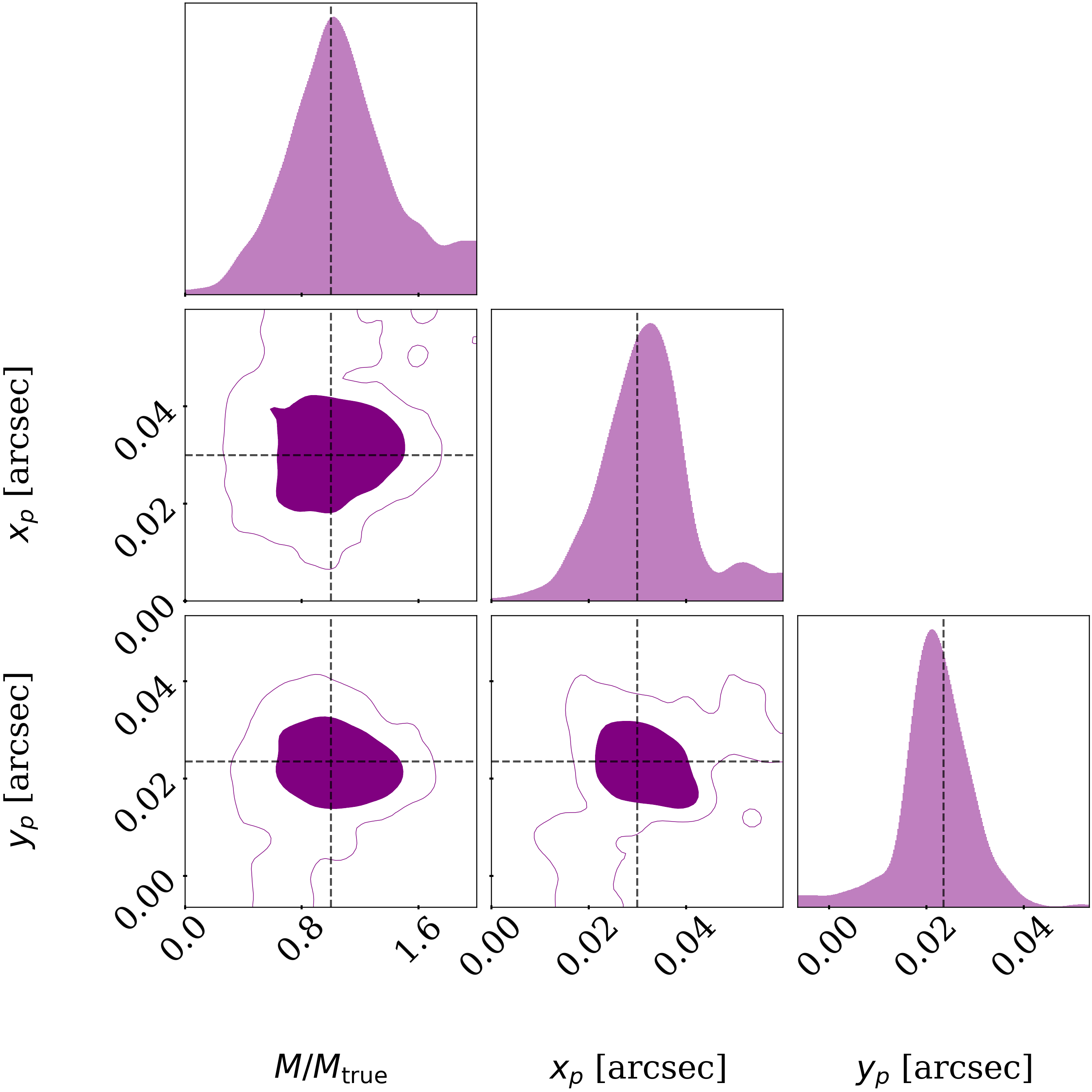}
    \caption{Posterior probability distributions for the parameters of a SMBH with mass $10^{7}\,\msun$ placed at the same optimal location as ``perturber 1".}
    \label{fig:bh_light}
\end{figure}

We also find that JWST-like data gives us a sensitivity to detect SMBHs down to $10^{7}\,\msun$. This is an order of magnitude lower in mass compared to detectable NFW perturbers. We show in Fig. \ref{fig:bh_light} the posterior probability distribution of the lighter black hole with mass $10^{7} \, \msun$. We see that we are at the threshold for a statistically significant detection with $M=0$ is excluded with a $3.1\sigma$ significance. If we increase our sensitivity by studying arcs with higher magnification, as we have explored in \cref{subsubsec:sensitiv}, we can push below $10^7\,\msun$.

\begin{figure}
    \centering
    \includegraphics[width=\linewidth]{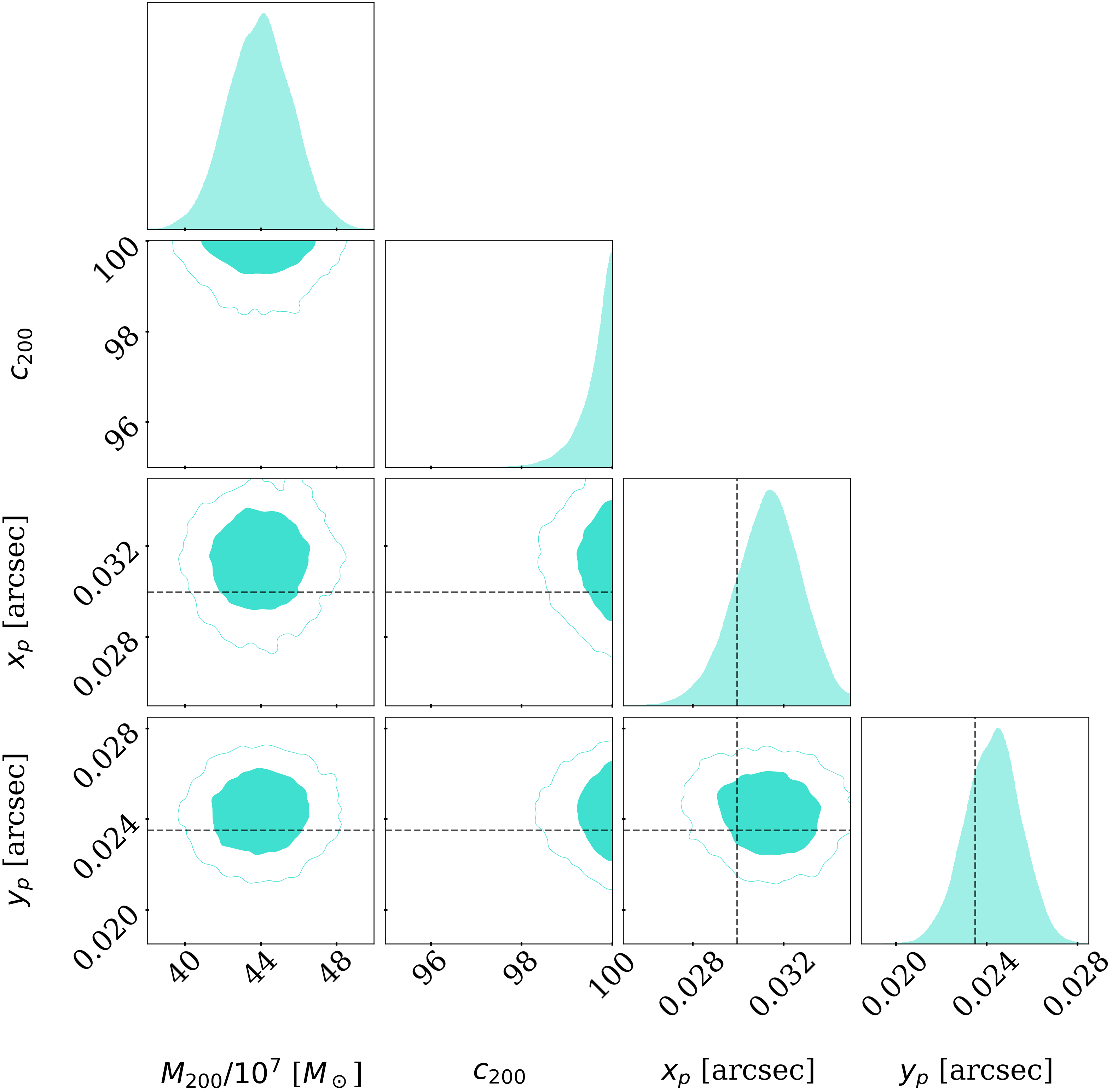}
    \caption{Posterior probability distributions for the NFW model parameters applied on a SMBH with mass $10^{8}\,\msun$ placed at the same optimal location as ``perturber 1".}
    \label{fig:bh_nfw}
\end{figure}

An important question is whether we can actually distinguish between a SMBH and an individual NFW subhalo. We test this by analyzing the mock data set with the $10^{8}\,\msun$ black hole using a NFW perturber lensing model given by Eq. \eqref{eq:nfw}. We show the posterior probability distribution of the NFW model parameters in Fig. \ref{fig:bh_nfw}. We chose a uniform prior range of $[1,100]$ for the concentration parameter $c_{200}$. This is motivated by the fact that the concentration of dark matter halos in our mass range is expected to be around 15 with a scatter of 0.2 decades \citep{mass_concen, 10.1093/mnras/stu483}. We see that the marginalized posteriors for the concentration $c_{200}$ of our SMBH hit the upper limit of the uniform prior. Comparing this to the concentration measurement of the NFW perturber with the same mass in Fig. \ref{fig:mock_post_lowmass}, where we recover the true concentration of $16$, we see that we can clearly differentiate SMBHs from NFW subhalos with masses $>10^{8}\,\msun$ by measuring their concentration.

%%%%%%%%%%%%%%%%%%%%%%%%%%%%%%%%%%%%%%%%%%%%%
\subsection{Recovery of Source Complexity}
\begin{figure*}
    \centering
    \includegraphics[width=\linewidth]{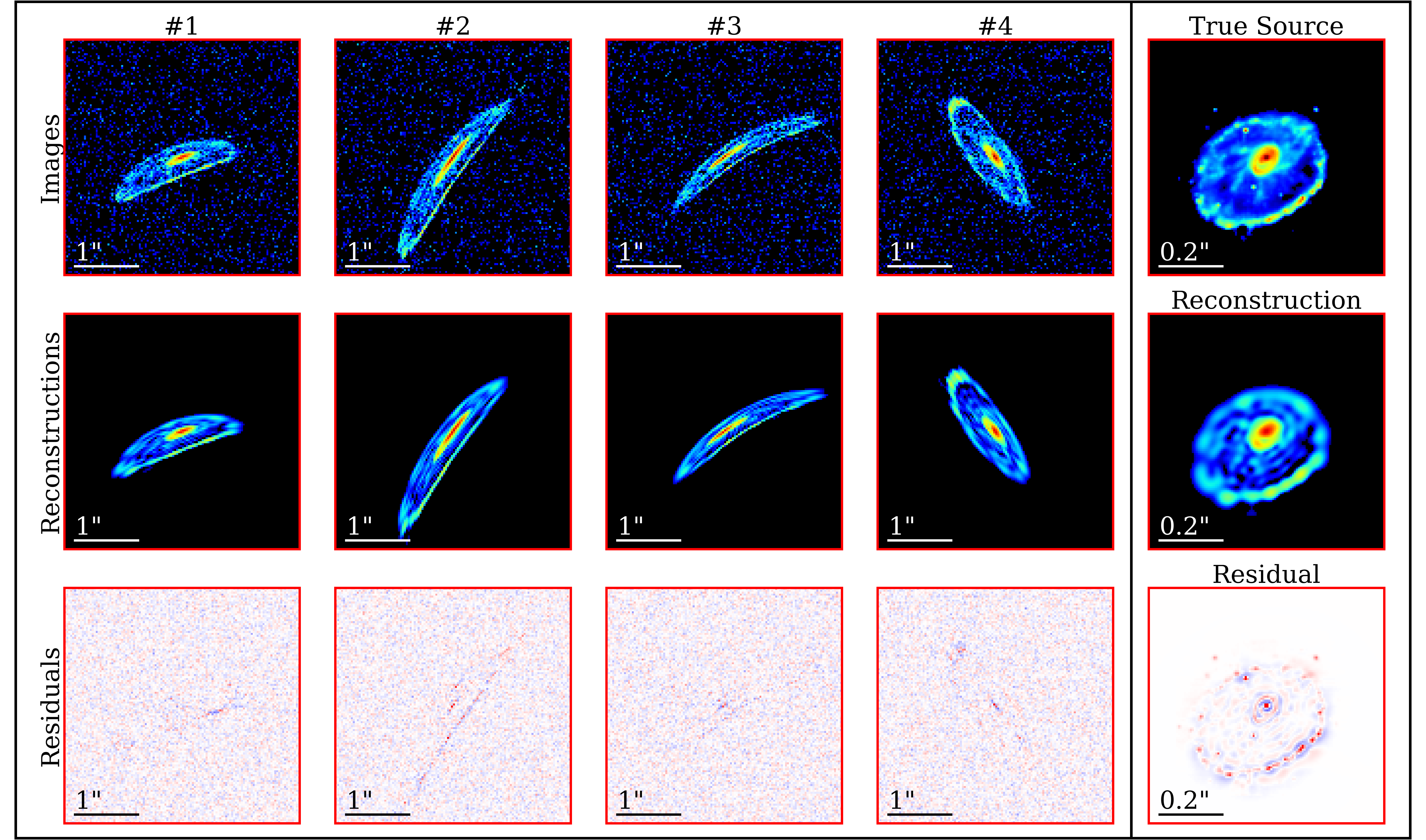}
    \caption{\textit{First row:} Four mock images created by lensing the image of the Cartwheel Galaxy with curved arc bases with different parameters. \textit{Second row:} The reconstructions of each image using the best fit of our source and lens model. \textit{Third row:} The normalized residuals between the data and the reconstruction. \textit{Fifth column:} The true source, its reconstruction, and its residual.}
    \label{fig:comp_source}
\end{figure*}

In addition to testing our ability to detect low-mass perturbers, we also study how our method handles a source light distribution that is much more complex than the one we analyzed in SMACS 0723 test case. In order to do this, we use the multiple images of the Cartwheel Galaxy (PGC 2248) taken by JWST as our mock source, since it is a galaxy with a very complex surface brightness distribution. We are motivated to use this complex shape as such a beaded, cartwheel shaped multiply imaged source has been detected in the HST image of the cluster lens CL0024+16 \citep{Colley_1996}. Here we use the higher-resolution JWST image of PGC 2248 to demonstrate the power of the curved arc basis method. We create a mock data set by lensing this mock source with four different curved arc bases, forming four different distorted images, as shown in the first row of Fig. \ref{fig:comp_source}. We then subsequently analyze these four images, simultaneously reconstructing the source and the angular deflections for each using the pipeline described in \cref{sec:smacs}, with the only difference being the number of images. Our model reconstruction and the residuals are shown in the second and third row of Fig. \ref{fig:comp_source}, respectively. We note that our method does not suffer from any problems handling a very complex source light distribution.

%%%%%%%%%%%%%%%%%%%%%%%%%%%%%%%%%%%%%%%%%%%%%
\section{Conclusions \& Discussion}\label{sec:conclusion}

Our work introduces a new method that models multiply imaged extended sources in cluster lenses using the curved arc basis. This method supplements the previous cluster lens analysis techniques that reconstruct mass distribution at larger scales by providing information on smaller scales. We capture the large-scale angular deflections caused by the smooth mass distribution of the cluster with an analytical angular deflection model. Any small-scale deflectors on top of this smooth deflection field result in residuals in the pixel-level model reconstructions in the lensed arcs, which can be used to detect the properties of these individual low-mass perturbers.
 
As we demonstrated in the previous section with tests on mock data, pixel-level modeling of gravitationally lensed arcs in cluster lenses with this novel method gives us the sensitivity to detect individual low mass perturbers down to $10^8\, \msun$ with JWST-level sensitivity. Our sensitivity to perturbers increases when the perturber is near a bright and sharp feature of the arc, when the perturber has a high concentration, or when the arc is stretched more dramatically by the cluster. For more massive perturbers with $10^9\, \msun$, we are sensitive to the properties of the density profile of the perturber, such as its concentration and ellipticity. For these more massive perturbers, we are also able to measure their redshift with an uncertainty of $~0.03$ allowing us to differentiate between line-of-sight halos and subhalos using only lensing data. These capabilities open up opportunities to use cluster lenses as a laboratory to study dark matter, as even the detection of a handful of low-mass subhalos (that may not host a visible baryonic component) and their density profiles can be used to differentiate between different dark matter models.
 
This method can also be used to probe the possibility of detecting compact dark sources like wandering SMBHs, which are predicted to exist in abundance in state-of-the-art cosmological simulations. Compared to NFW perturbers, we are more sensitive to black holes as they are point masses, which results in stronger angular deflections for a given mass. We find that, with JWST-level sensitivity, we are able to detect SMBHs down to $10^7\, \msun$ near observed elongated cluster arcs. Additionally, we show that we can distinguish between a SMBH and an NFW perturber by measuring the concentration of a given deflector. Gravitational lensing offers the only method to detect these black holes if they are not actively accreting, providing the unique possibility to gain insights into the origin and evolution of galaxies.

Our method works best when the arcs are distinct images of the source, as for these situations the validity of the curved arc basis would cover the entire image. When the arcs are the result of multiple image mergers, a single curved arc basis would not be able to capture the angular deflections across the critical curve. One possible way to circumvent this problem is having a separate curved arc basis for each side of the critical curve and masking the regions where the approximation fails. One needs to consider a different basis for the angular deflections near the critical curve for the images that are folds and cusps, which we leave for future work.

%=================================
\vspace{0.5cm}
\subsection*{Acknowledgments} 

CD and ACS are partially supported by the Department of Energy (DOE) Grant No. DE-SC0020223. 
SB is partially supported by NASA through grant number HST-GO 17199 from the Space Telescope Science Institute.
PN acknowledges support from the Gordon and Betty Moore Foundation and the John Templeton Foundation that fund the Black Hole Initiative where she serves as one of the PIs. ACS would like to thank Arthur Tsang for useful discussions and comments.

%%%%%%%%%%%%%%%%%%%%%%%%%%%%%%%%%%%%%%%%%%%%%
\subsection*{Data Availability} 
The system analyzed in this work is the JWST image of the galaxy cluster SMACS J0723.3-7327. The data of this system are publicly available on \url{https://mast.stsci.edu/portal/Mashup/Clients/Mast/Portal.html}. The code used for our analysis is available at \url{https://github.com/acagansengul/local_cluster_lensing.git}.

%%%%%%%%%%%%%%%%%%%%%%%%%%%%%%%%%%%%%%%%%%%%%
%\begin{appendix}

%%%%%%%%%%%%%%%%%%%%%%%%%%%%%%%%%%%%%%%%%%%%%
%\section{Add appendix}
%\label{sec:apx1}

%\end{appendix}

\newpage
\appendix

\section{Perturbers as Effective Power Law Profiles}
In this appendix, we present a number of tests with realistic mock images. Same as the analysis in \cref{sec:JWSTsims}, data properties such as PSF, exposure, noise, and resolution are all set to be identical to the JWST image of the cluster lens SMACS 0723 analyzed earlier. Our goal here is to investigate how well our method performs when the mass profile of the perturber model does not match that of the true perturber. Same as in \cref{sec:JWSTsims}, we use the best fits of the smooth-lens model parameters that we obtained from our analysis of SMACS 0723, as well as the source reconstruction, to create multiple mock data sets. Each of the data sets consists of three images of the same source. These three images are shown in the top rows of Figs. \ref{fig:slope23}, \ref{fig:slope20}, \ref{fig:slope17}, and \ref{fig:slope14}. In each data set, we have placed a perturber near one of the bright images of the background source. This perturber is placed in the same location as "perturber 1" shown in Fig. \ref{fig:mock}. The convergence of the perturber in the mock images is given by a power-law (PL) profile:
\begin{equation}\label{eq:epl}
    \kappa(x, y) = \frac{3-\gamma}{2} \left(\frac{\theta_{E}}{\sqrt{x^2 + y^2}} \right)^{\gamma-1},
\end{equation}
where $\theta_{E}$ is the Einstein radius and $\gamma$ is the power-law slope. The isothermal mass distribution corresponds to a slope value of $\gamma = 2$. We made four datasets, with the value of the power-law slope of the perturber set to $\gamma = 2.3,2.0,1.7,\,\text{and}\, 1.4$. As we vary the slope, we also change $\theta_{E}$ to keep the mass within $0.1''$ constant to a value of $10^9\,\msun$ as we calculate that the region of maximum observability \cite{effective_power_law} for this system is between $0.05''$ and $0.20''$.

\subsection*{Recovering the True Model Parameters}
\begin{figure}
    \centering
    \includegraphics[width=\linewidth]{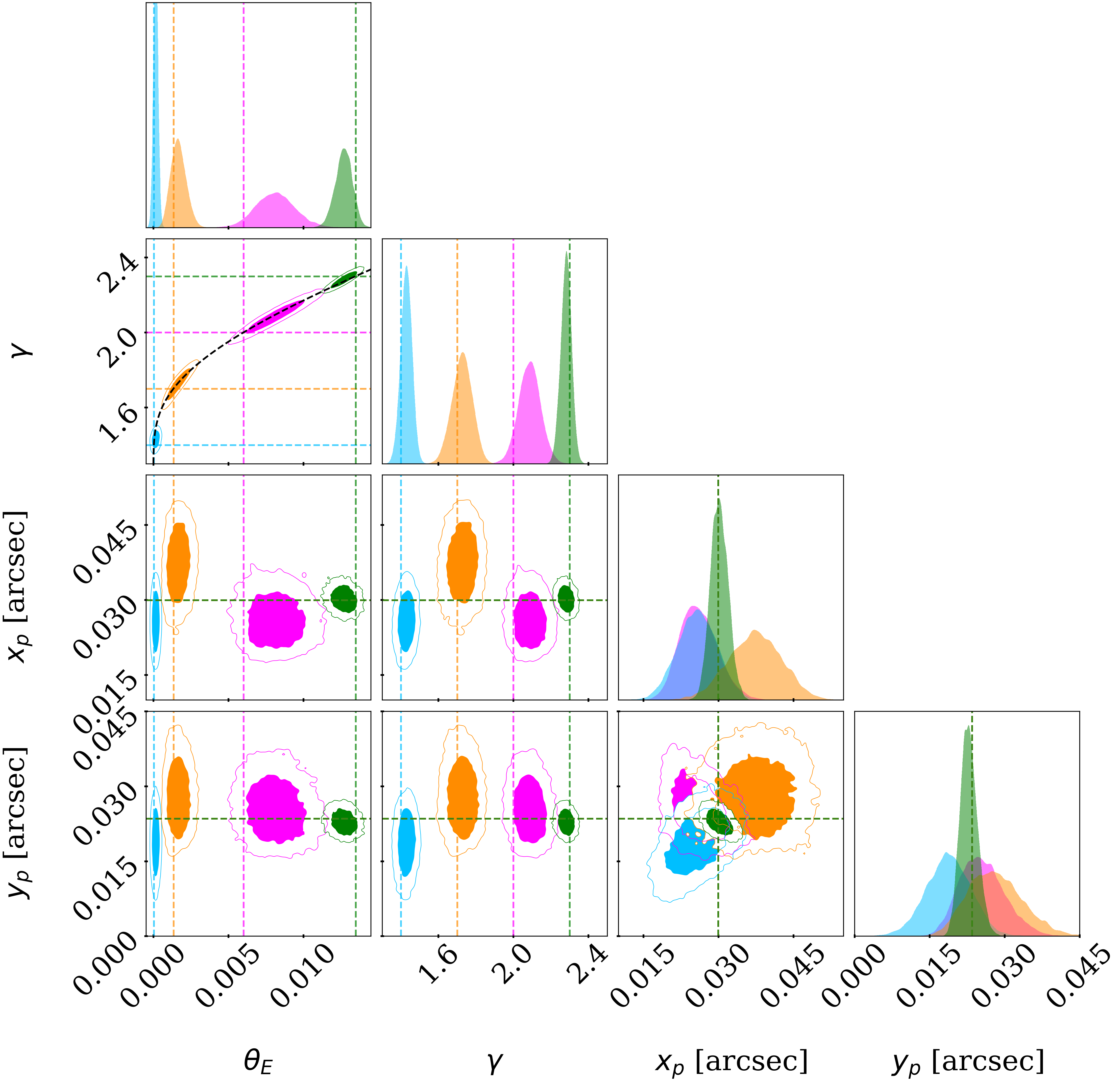}
    \caption{Posterior probability distributions for the power-law model parameters applied on power-law perturbers placed at the same location as ``perturber 1". The true parameter values are shown as colored dashed lines. The black line shows the pairs of values for the slope and the Einstein radius for the power-law profile that have the same total projected mass within 0.2$''$ set to $10^9\,\msun$.}
    \label{fig:pl_pltrue}
\end{figure}

We first analyze these mock datasets with the same lens model that was used to create them. We present the posterior probability distributions of the perturber parameters in Fig. \ref{fig:pl_pltrue}, where see that the parameter posteriors are consistent with their true values. The correlation between $\theta_E$ and $\gamma$ lies along the line that shows the total projected mass of $10^9\,\msun$ within $0.1''$ (dashed black line in Fig. \ref{fig:pl_pltrue}). We also see that the positions of the perturbers with the power-law slopes that are isothermal or shallower than isothermal ($\gamma \leq 2.0$) are constrained more poorly compared to that of the perturber with a steeper power-law profile ($\gamma = 2.3$). This is expected as a steep power-law slope results in stronger angular deflections closer to the perturber's center.

\subsection*{Analyzing the Dataset with an NFW Model}
\begin{figure}
    \centering
    \includegraphics[width=\linewidth]{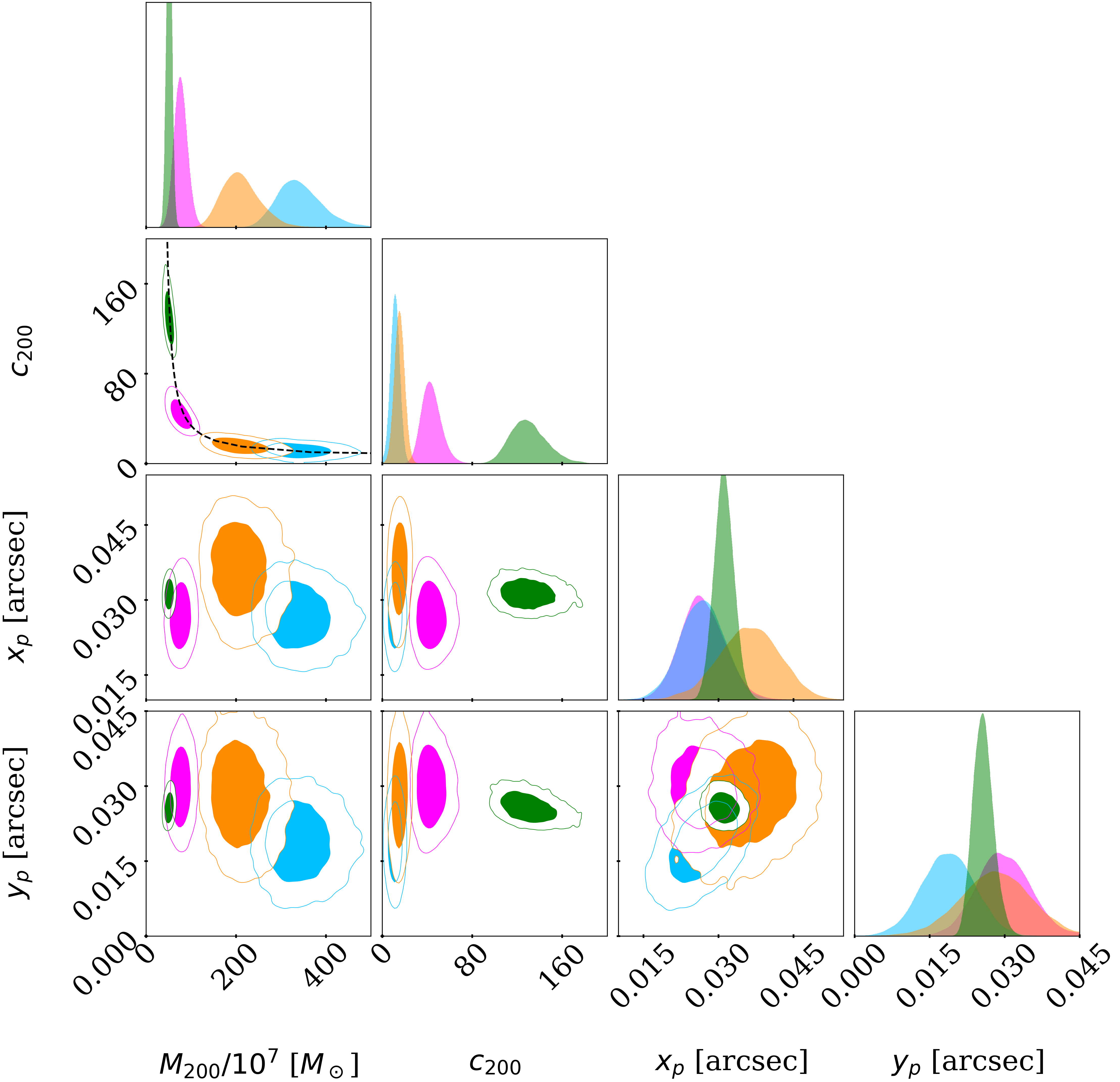}
    \caption{Posterior probability distributions for the NFW parameters applied on power-law perturbers placed at the same location as ``perturber 1". The true parameter values are shown as colored dashed lines. The dashed black line shows the pairs of values for the concentration and mass for the NFW profile that have the same total projected mass within 0.2$''$ set to $10^9\,\msun$.}
    \label{fig:pl_plnfw}
\end{figure}

We also analyze the mock datasets with a model that has an NFW perturber (the profile given in Eq. \eqref{eq:nfw}). We present the posterior probability distributions of the perturber parameters in Fig. \ref{fig:pl_plnfw}, where we see that the NFW model gives slightly larger uncertainties for the perturber position compared to the power-law model, but it does not bias the position measurement. We find that the difference between the goodness-of-fit of NFW profle and true power-law model is statistically insignificant for all of the four perturbers. We also find that the total projected mass within $0.1''$ that is inferred from the NFW model is consistent with that of the true power-law profile. Moreover, the correlation between $M_{200}$ and $c_{200}$ also lies along the line that shows the total projected mass of $10^9\,\msun$ within $0.1''$ (dashed black line in Fig. \ref{fig:pl_plnfw}). We see that there is a correspondance between the power-law slope and the concentration of the equivalent NFW perturber, with a higher concentration implying a steeper slope.

\subsection*{Image and Source Reconstructions}
\begin{figure*}
    \centering
    \includegraphics[width=0.8\linewidth]{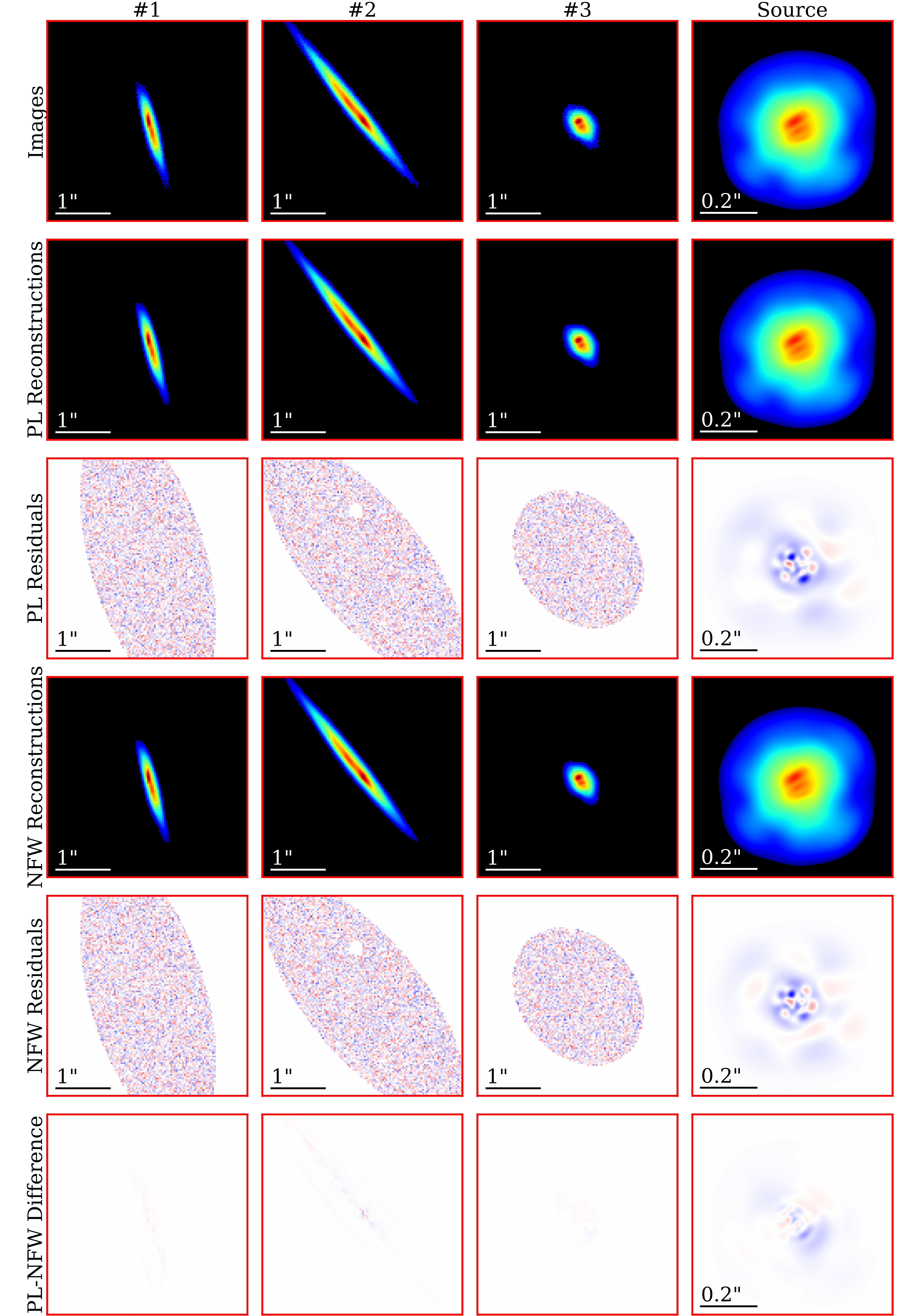}
    \caption{\textit{First row:} Three mock images created with a perturber with a power-law slope of $\gamma = 2.3$. The fourth box shows the true source. \textit{Second row:} The reconstructions of each image using the best fit of our source and lens model, where we use a power-law model for the perturber. The fourth box shows the model reconstruction of the source. \textit{Third row:} The normalized residuals between the data and the power-law reconstruction. The fourth box shows the difference between the true source and the power-law reconstruction.  \textit{Fourth Row:} The reconstructions of each image using the best fit of our source and lens model where we use an NFW model for the perturber. The fourth box shows the model reconstruction of the source. \textit{Fifth row:} The normalized residuals between the data and the NFW reconstruction. The fourth box shows the difference between the true source and the NFW reconstruction. \textit{Sixth row:} The normalized residuals between the power-law and the NFW reconstruction. The fourth box shows the difference between the power-law and the NFW reconstruction of the source. The width of the color scale in the image residuals correspond to 3$\sigma$ in pixel errors. The width of the color scale in the source residuals corresponds to a 1\% difference between the source reconstructions compared to the brightest pixel.}
    \label{fig:slope23}
\end{figure*}

\begin{figure*}
    \centering
    \includegraphics[width=0.8\linewidth]{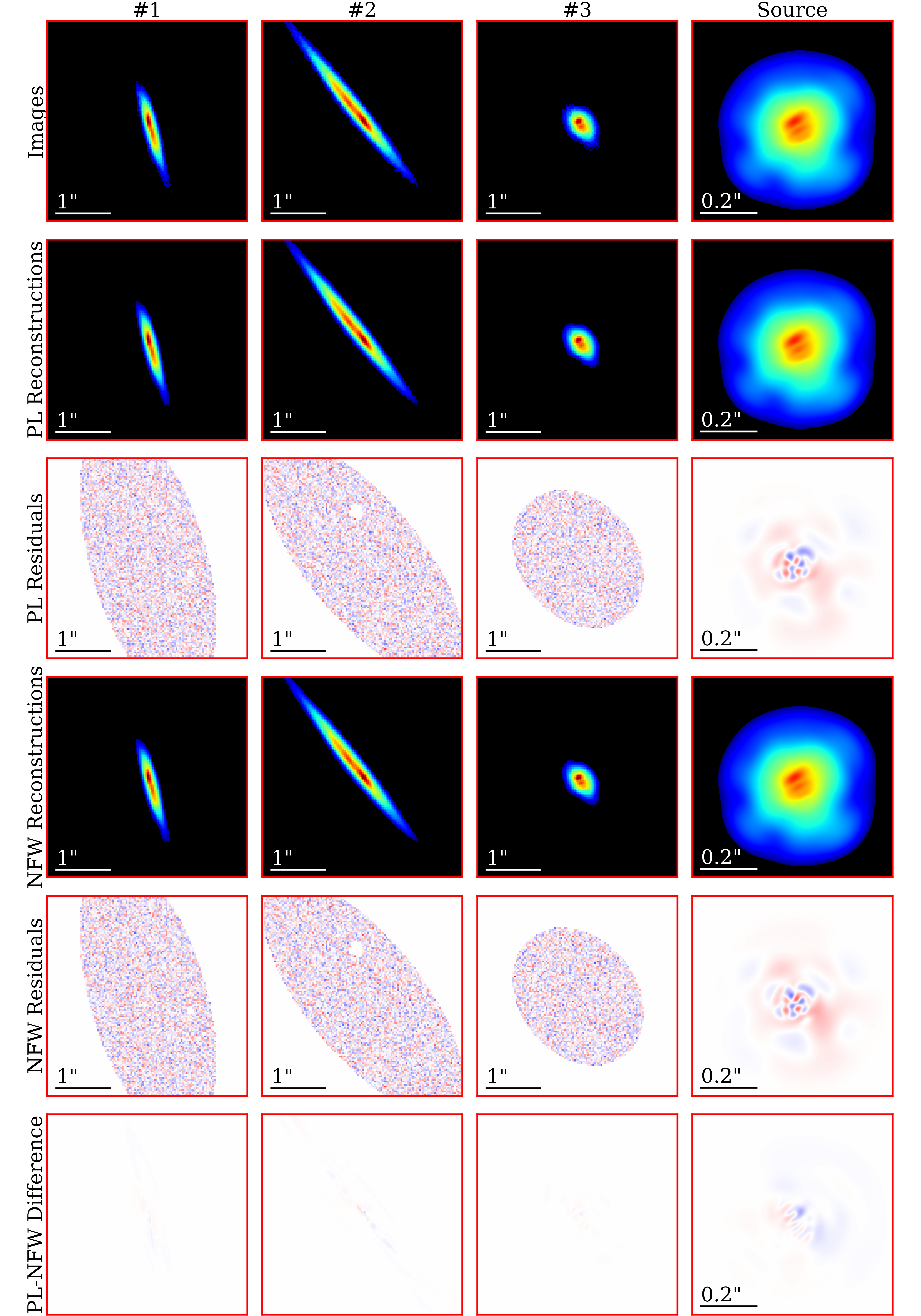}
    \caption{Same as Fig. \ref{fig:slope23} but for mock images created with a perturber with a power-law slope of $\gamma = 2.0$.}
    \label{fig:slope20}
\end{figure*}

\begin{figure*}
    \centering
    \includegraphics[width=0.8\linewidth]{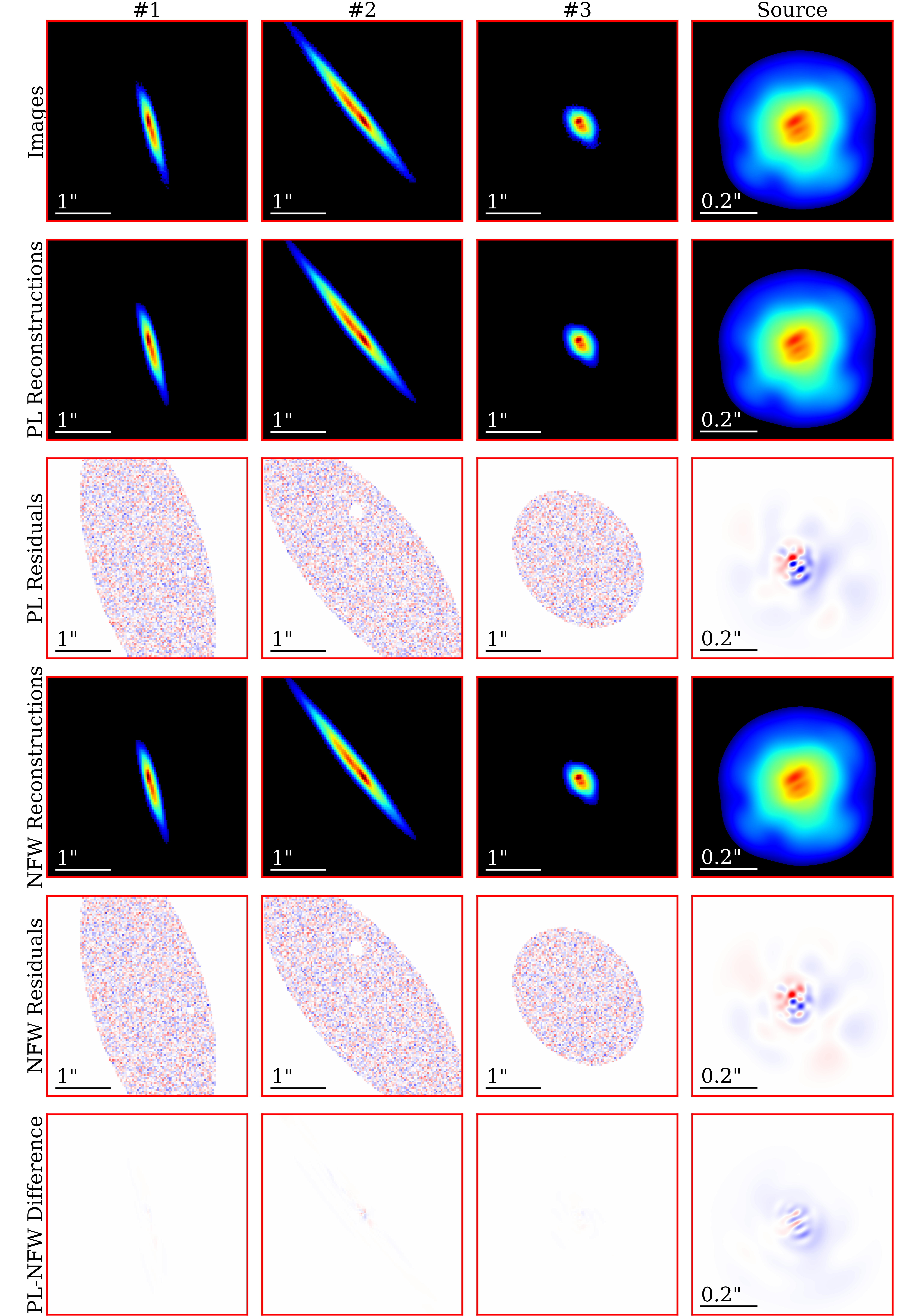}
    \caption{Same as Fig. \ref{fig:slope23} but for mock images created with a perturber with a power-law slope of $\gamma = 1.7$.}
    \label{fig:slope17}
\end{figure*}

\begin{figure*}
    \centering
    \includegraphics[width=0.8\linewidth]{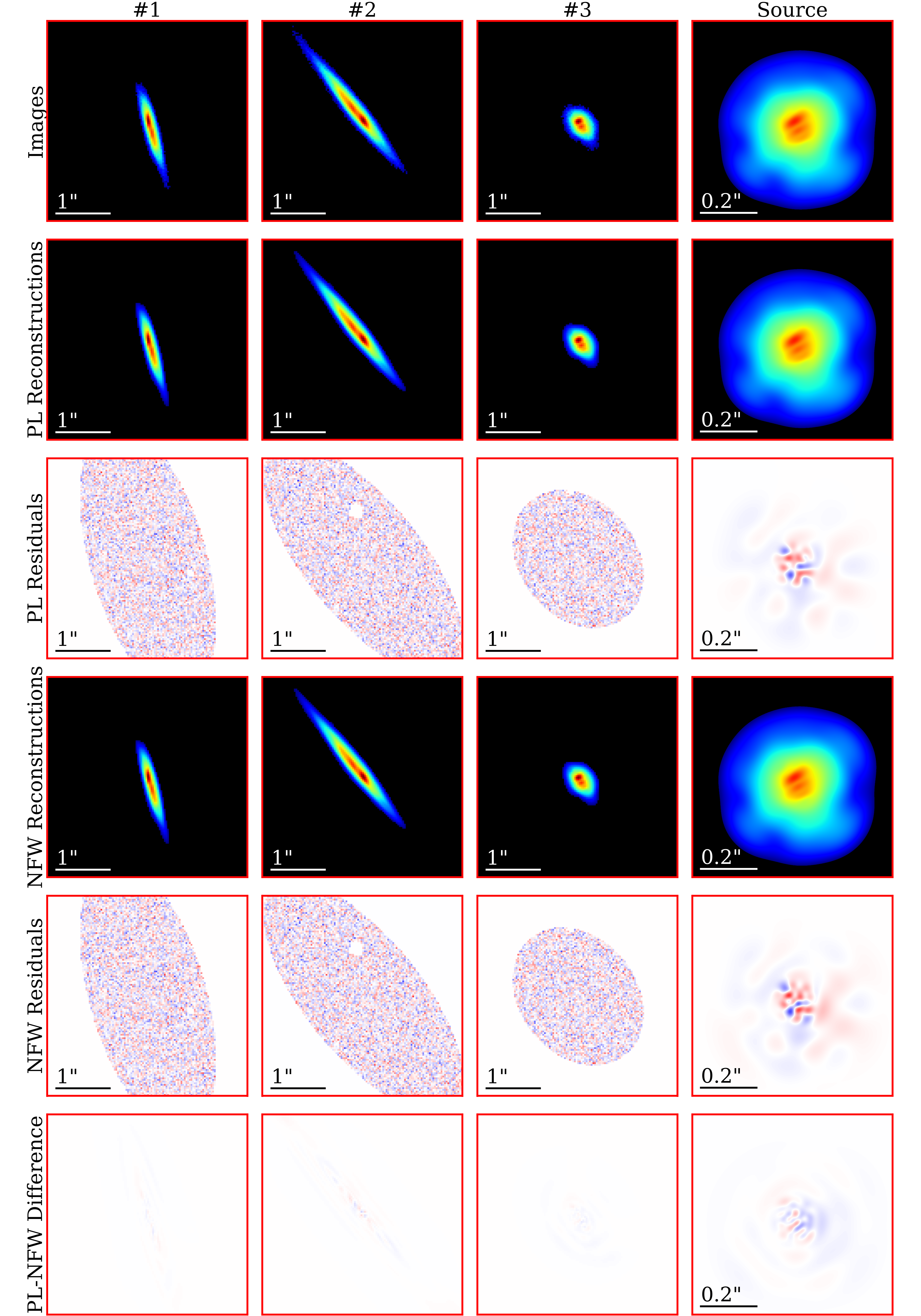}
    \caption{Same as Fig. \ref{fig:slope23} but for mock images created with a perturber with a power-law slope of $\gamma = 1.4$.}
    \label{fig:slope14}
\end{figure*}

In the fourth column of Figs. \ref{fig:slope23}, \ref{fig:slope20}, \ref{fig:slope17}, and \ref{fig:slope14} we show the source reconstructions of the power-law model and the NFW model, along with the true source and their differences. All of these reconstructions were made with the best fits of each model. We see no significant difference between the source reconstructions of the power-law and NFW models. We are capable of accurately reconstructing the source even if the mass profile of our model for the perturber differs from the actual mass profile of the perturber. The difference between the surface brightness distribution of the true source and the reconstructed source is always less than 1\% of the brightest pixel of the source.

\newpage
\hspace{1cm}
\newpage
\hspace{1cm}
\newpage
\hspace{1cm}
\newpage
\hspace{1cm}
\newpage
\hspace{1cm}
\newpage
\hspace{1cm}
\newpage

\bibliographystyle{mnras}
\bibliography{main} 

\begin{thebibliography}{}
\makeatletter
\relax
\def\mn@urlcharsother{\let\do\@makeother \do\$\do\&\do\#\do\^\do\_\do\%\do\~}
\def\mn@doi{\begingroup\mn@urlcharsother \@ifnextchar [ {\mn@doi@}
  {\mn@doi@[]}}
\def\mn@doi@[#1]#2{\def\@tempa{#1}\ifx\@tempa\@empty \href
  {http://dx.doi.org/#2} {doi:#2}\else \href {http://dx.doi.org/#2} {#1}\fi
  \endgroup}
\def\mn@eprint#1#2{\mn@eprint@#1:#2::\@nil}
\def\mn@eprint@arXiv#1{\href {http://arxiv.org/abs/#1} {{\tt arXiv:#1}}}
\def\mn@eprint@dblp#1{\href {http://dblp.uni-trier.de/rec/bibtex/#1.xml}
  {dblp:#1}}
\def\mn@eprint@#1:#2:#3:#4\@nil{\def\@tempa {#1}\def\@tempb {#2}\def\@tempc
  {#3}\ifx \@tempc \@empty \let \@tempc \@tempb \let \@tempb \@tempa \fi \ifx
  \@tempb \@empty \def\@tempb {arXiv}\fi \@ifundefined
  {mn@eprint@\@tempb}{\@tempb:\@tempc}{\expandafter \expandafter \csname
  mn@eprint@\@tempb\endcsname \expandafter{\@tempc}}}

\bibitem[\protect\citeauthoryear{Abbott et~al.,}{Abbott
  et~al.}{2018}]{des_matter_power}
Abbott T. M.~C.,  et~al., 2018, \mn@doi [Phys. Rev. D]
  {10.1103/PhysRevD.98.043526}, 98, 043526

\bibitem[\protect\citeauthoryear{{Abolfathi} et~al.,}{{Abolfathi}
  et~al.}{2018}]{lyman_alpha}
{Abolfathi} B.,  et~al., 2018, \mn@doi [\apjs] {10.3847/1538-4365/aa9e8a},
  \href {https://ui.adsabs.harvard.edu/abs/2018ApJS..235...42A} {235, 42}

\bibitem[\protect\citeauthoryear{Bergamini et~al.,}{Bergamini
  et~al.}{2019}]{Bergamini2019}
Bergamini P.,  et~al., 2019, \mn@doi [A\&A] {10.1051/0004-6361/201935974}, 631,
  A130

\bibitem[\protect\citeauthoryear{Bergamini et~al.,}{Bergamini
  et~al.}{2021}]{Bergamini2021}
Bergamini P.,  et~al., 2021, \mn@doi [A\&A] {10.1051/0004-6361/202039564}, 645,
  A140

\bibitem[\protect\citeauthoryear{{Birrer}}{{Birrer}}{2021}]{curved_arc_theory}
{Birrer} S.,  2021, \mn@doi [\apj] {10.3847/1538-4357/ac1108}, \href
  {https://ui.adsabs.harvard.edu/abs/2021ApJ...919...38B} {919, 38}

\bibitem[\protect\citeauthoryear{{Birrer} \& {Amara}}{{Birrer} \&
  {Amara}}{2018}]{lenstronomy1}
{Birrer} S.,  {Amara} A.,  2018, \mn@doi [Physics of the Dark Universe]
  {10.1016/j.dark.2018.11.002}, \href
  {https://ui.adsabs.harvard.edu/abs/2018PDU....22..189B} {22, 189}

\bibitem[\protect\citeauthoryear{{Birrer}, {Amara}  \& {Refregier}}{{Birrer}
  et~al.}{2015}]{lenstronomy_shape}
{Birrer} S.,  {Amara} A.,   {Refregier} A.,  2015, \mn@doi [\apj]
  {10.1088/0004-637X/813/2/102}, \href
  {https://ui.adsabs.harvard.edu/abs/2015ApJ...813..102B} {813, 102}

\bibitem[\protect\citeauthoryear{Birrer et~al.}{Birrer
  et~al.}{2019}]{Birrer:2018vtm}
Birrer S.,  et~al., 2019, \mn@doi [\mnras] {10.1093/mnras/stz200}, 484, 4726

\bibitem[\protect\citeauthoryear{Birrer et~al.,}{Birrer
  et~al.}{2021}]{lenstronomy2}
Birrer S.,  et~al., 2021, \mn@doi [Journal of Open Source Software]
  {10.21105/joss.03283}, 6, 3283

\bibitem[\protect\citeauthoryear{{Bonamigo}, {Despali}, {Limousin}, {Angulo},
  {Giocoli}  \& {Soucail}}{{Bonamigo} et~al.}{2015}]{2015MNRAS.449.3171B}
{Bonamigo} M.,  {Despali} G.,  {Limousin} M.,  {Angulo} R.,  {Giocoli} C.,
  {Soucail} G.,  2015, \mn@doi [\mnras] {10.1093/mnras/stv417}, \href
  {https://ui.adsabs.harvard.edu/abs/2015MNRAS.449.3171B} {449, 3171}

\bibitem[\protect\citeauthoryear{{Caminha}, {Suyu}, {Mercurio}, {Brammer},
  {Bergamini}, {Acebron}  \& {Vanzella}}{{Caminha}
  et~al.}{2022}]{smacs_jwst_model1}
{Caminha} G.~B.,  {Suyu} S.~H.,  {Mercurio} A.,  {Brammer} G.,  {Bergamini} P.,
   {Acebron} A.,   {Vanzella} E.,  2022, \mn@doi [\aap]
  {10.1051/0004-6361/202244517}, \href
  {https://ui.adsabs.harvard.edu/abs/2022A&A...666L...9C} {666, L9}

\bibitem[\protect\citeauthoryear{{Colley}, {Tyson}  \& {Turner}}{{Colley}
  et~al.}{1996}]{Colley_1996}
{Colley} W.~N.,  {Tyson} J.~A.,   {Turner} E.~L.,  1996, \mn@doi [\apjl]
  {10.1086/310015}, \href
  {https://ui.adsabs.harvard.edu/abs/1996ApJ...461L..83C} {461, L83}

\bibitem[\protect\citeauthoryear{{D'Aloisio} \& {Natarajan}}{{D'Aloisio} \&
  {Natarajan}}{2011}]{los_cluster}
{D'Aloisio} A.,  {Natarajan} P.,  2011, \mn@doi [\mnras]
  {10.1111/j.1365-2966.2010.17795.x}, \href
  {https://ui.adsabs.harvard.edu/abs/2011MNRAS.411.1628D} {411, 1628}

\bibitem[\protect\citeauthoryear{{De Propris}, {Bremer}  \& {Phillipps}}{{De
  Propris} et~al.}{2018}]{DePropis_2018}
{De Propris} R.,  {Bremer} M.~N.,   {Phillipps} S.,  2018, \mn@doi [\aap]
  {10.1051/0004-6361/201833630}, \href
  {https://ui.adsabs.harvard.edu/abs/2018A&A...618A.180D} {618, A180}

\bibitem[\protect\citeauthoryear{Diego, Pascale, Kavanagh, Kelly, Dai, Frye  \&
  Broadhurst}{Diego et~al.}{2022}]{Diego2022}
Diego J.~M.,  Pascale M.,  Kavanagh B.~J.,  Kelly P.,  Dai L.,  Frye B.,
  Broadhurst T.,  2022, \mn@doi [A\&A] {10.1051/0004-6361/202243605}, 665, A134

\bibitem[\protect\citeauthoryear{{Dutton} \& {Macci{\`o}}}{{Dutton} \&
  {Macci{\`o}}}{2014}]{mass_concen}
{Dutton} A.~A.,  {Macci{\`o}} A.~V.,  2014, \mn@doi [\mnras]
  {10.1093/mnras/stu742}, \href
  {https://ui.adsabs.harvard.edu/abs/2014MNRAS.441.3359D} {441, 3359}

\bibitem[\protect\citeauthoryear{{Ebeling}, {Edge}  \& {Henry}}{{Ebeling}
  et~al.}{2001}]{smacs_discovery}
{Ebeling} H.,  {Edge} A.~C.,   {Henry} J.~P.,  2001, \mn@doi [\apj]
  {10.1086/320958}, \href
  {https://ui.adsabs.harvard.edu/abs/2001ApJ...553..668E} {553, 668}

\bibitem[\protect\citeauthoryear{Gil-Marín et~al.,}{Gil-Marín
  et~al.}{2016}]{boss_matter_power}
Gil-Marín H.,  et~al., 2016, \mn@doi [\mnras] {10.1093/mnras/stw1096}, 460,
  4188

\bibitem[\protect\citeauthoryear{{Golse} \& {Kneib}}{{Golse} \&
  {Kneib}}{2002}]{elliptical_nfw}
{Golse} G.,  {Kneib} J.~P.,  2002, \mn@doi [\aap] {10.1051/0004-6361:20020639},
  \href {https://ui.adsabs.harvard.edu/abs/2002A&A...390..821G} {390, 821}

\bibitem[\protect\citeauthoryear{{Golubchik}, {Furtak}, {Meena}  \&
  {Zitrin}}{{Golubchik} et~al.}{2022}]{smacs_hst_model1}
{Golubchik} M.,  {Furtak} L.~J.,  {Meena} A.~K.,   {Zitrin} A.,  2022, \mn@doi
  [\apj] {10.3847/1538-4357/ac8ff1}, \href
  {https://ui.adsabs.harvard.edu/abs/2022ApJ...938...14G} {938, 14}

\bibitem[\protect\citeauthoryear{{Jing} \& {Suto}}{{Jing} \&
  {Suto}}{2002}]{2002ApJ...574..538J}
{Jing} Y.~P.,  {Suto} Y.,  2002, \mn@doi [\apj] {10.1086/341065}, \href
  {https://ui.adsabs.harvard.edu/abs/2002ApJ...574..538J} {574, 538}

\bibitem[\protect\citeauthoryear{{Kneib} \& {Natarajan}}{{Kneib} \&
  {Natarajan}}{2011}]{cluster_lens_review}
{Kneib} J.-P.,  {Natarajan} P.,  2011, \mn@doi [\aapr]
  {10.1007/s00159-011-0047-3}, \href
  {https://ui.adsabs.harvard.edu/abs/2011A&ARv..19...47K} {19, 47}

\bibitem[\protect\citeauthoryear{{Li}, {Frenk}, {Cole}, {Wang}  \& {Gao}}{{Li}
  et~al.}{2017}]{mass_redshift_degen}
{Li} R.,  {Frenk} C.~S.,  {Cole} S.,  {Wang} Q.,   {Gao} L.,  2017, \mn@doi
  [\mnras] {10.1093/mnras/stx554}, \href
  {https://ui.adsabs.harvard.edu/abs/2017MNRAS.468.1426L} {468, 1426}

\bibitem[\protect\citeauthoryear{Ludlow, Navarro, Angulo, Boylan-Kolchin,
  Springel, Frenk  \& White}{Ludlow et~al.}{2014}]{10.1093/mnras/stu483}
Ludlow A.~D.,  Navarro J.~F.,  Angulo R.~E.,  Boylan-Kolchin M.,  Springel V.,
  Frenk C.,   White S. D.~M.,  2014, \mn@doi [\mnras] {10.1093/mnras/stu483},
  441, 378

\bibitem[\protect\citeauthoryear{{Mahler} et~al.,}{{Mahler}
  et~al.}{2022a}]{smacs_redshifts}
{Mahler} G.,  et~al., 2022a, arXiv e-prints, \href
  {https://ui.adsabs.harvard.edu/abs/2022arXiv220707101M} {p. arXiv:2207.07101}

\bibitem[\protect\citeauthoryear{Mahler, Natarajan, Jauzac  \& Richard}{Mahler
  et~al.}{2022b}]{wandering_bh_lensing}
Mahler G.,  Natarajan P.,  Jauzac M.,   Richard J.,  2022b, \mn@doi [\mnras]
  {10.1093/mnras/stac3098}, 518, 54

\bibitem[\protect\citeauthoryear{{Meneghetti} et~al.,}{{Meneghetti}
  et~al.}{2017}]{Frontier_Fields}
{Meneghetti} M.,  et~al., 2017, \mn@doi [\mnras] {10.1093/mnras/stx2064}, \href
  {https://ui.adsabs.harvard.edu/abs/2017MNRAS.472.3177M} {472, 3177}

\bibitem[\protect\citeauthoryear{Meneghetti et~al.,}{Meneghetti
  et~al.}{2020}]{meneghetti2020}
Meneghetti M.,  et~al., 2020, \mn@doi [Science] {10.1126/science.aax5164}, 369,
  1347

\bibitem[\protect\citeauthoryear{Meneghetti et~al.,}{Meneghetti
  et~al.}{2022}]{meneghetti2022}
Meneghetti M.,  et~al., 2022, \mn@doi [A\&A] {10.1051/0004-6361/202243779},
  668, A188

\bibitem[\protect\citeauthoryear{Monna et~al.,}{Monna
  et~al.}{2016}]{monna2016a}
Monna A.,  et~al., 2016, \mn@doi [Monthly Notices of the Royal Astronomical
  Society] {10.1093/mnras/stw3048}, 465, 4589

\bibitem[\protect\citeauthoryear{Monna et~al.,}{Monna et~al.}{2017}]{monna2017}
Monna A.,  et~al., 2017, \mn@doi [Monthly Notices of the Royal Astronomical
  Society] {10.1093/mnras/stx015}, 466, 4094

\bibitem[\protect\citeauthoryear{Natarajan \& Kneib}{Natarajan \&
  Kneib}{1997}]{Natarajan_1997}
Natarajan P.,  Kneib J.-P.,  1997, \mn@doi [\mnras] {10.1093/mnras/287.4.833},
  287, 833

\bibitem[\protect\citeauthoryear{{Natarajan}, {De Lucia}  \&
  {Springel}}{{Natarajan} et~al.}{2007}]{mass_range}
{Natarajan} P.,  {De Lucia} G.,   {Springel} V.,  2007, \mn@doi [\mnras]
  {10.1111/j.1365-2966.2007.11399.x}, \href
  {https://ui.adsabs.harvard.edu/abs/2007MNRAS.376..180N} {376, 180}

\bibitem[\protect\citeauthoryear{{Natarajan} et~al.,}{{Natarajan}
  et~al.}{2017}]{Natarajan+2017}
{Natarajan} P.,  et~al., 2017, \mn@doi [\mnras] {10.1093/mnras/stw3385}, \href
  {https://ui.adsabs.harvard.edu/abs/2017MNRAS.468.1962N} {468, 1962}

\bibitem[\protect\citeauthoryear{{Navarro}, {Frenk}  \& {White}}{{Navarro}
  et~al.}{1996}]{NFW}
{Navarro} J.~F.,  {Frenk} C.~S.,   {White} S. D.~M.,  1996, \mn@doi [\apj]
  {10.1086/177173}, \href
  {https://ui.adsabs.harvard.edu/abs/1996ApJ...462..563N} {462, 563}

\bibitem[\protect\citeauthoryear{{Navarro}, {Frenk}  \& {White}}{{Navarro}
  et~al.}{1997}]{NFW2}
{Navarro} J.~F.,  {Frenk} C.~S.,   {White} S. D.~M.,  1997, \mn@doi [\apj]
  {10.1086/304888}, \href
  {https://ui.adsabs.harvard.edu/abs/1997ApJ...490..493N} {490, 493}

\bibitem[\protect\citeauthoryear{{Newman}, {Treu}, {Ellis}, {Sand}, {Nipoti},
  {Richard}  \& {Jullo}}{{Newman} et~al.}{2013}]{cnfw}
{Newman} A.~B.,  {Treu} T.,  {Ellis} R.~S.,  {Sand} D.~J.,  {Nipoti} C.,
  {Richard} J.,   {Jullo} E.,  2013, \mn@doi [\apj]
  {10.1088/0004-637X/765/1/24}, \href
  {https://ui.adsabs.harvard.edu/abs/2013ApJ...765...24N} {765, 24}

\bibitem[\protect\citeauthoryear{Parry et~al.,}{Parry et~al.}{2016}]{parry2016}
Parry W.~G.,  et~al., 2016, \mn@doi [Monthly Notices of the Royal Astronomical
  Society] {10.1093/mnras/stw298}, 458, 1493

\bibitem[\protect\citeauthoryear{Pascale et~al.,}{Pascale
  et~al.}{2022}]{Pascale_2022}
Pascale M.,  et~al., 2022, \mn@doi [The Astrophysical Journal Letters]
  {10.3847/2041-8213/ac9316}, 938, L6

\bibitem[\protect\citeauthoryear{Pignataro et~al.,}{Pignataro
  et~al.}{2021}]{Pignataro2021}
Pignataro G.~V.,  et~al., 2021, \mn@doi [A\&A] {10.1051/0004-6361/202141586},
  655, A81

\bibitem[\protect\citeauthoryear{{Planck Collaboration} et~al.,}{{Planck
  Collaboration} et~al.}{2020}]{planck2018}
{Planck Collaboration} et~al., 2020, \mn@doi [\aap]
  {10.1051/0004-6361/201833910}, \href
  {https://ui.adsabs.harvard.edu/abs/2020A&A...641A...6P} {641, A6}

\bibitem[\protect\citeauthoryear{{Refregier}}{{Refregier}}{2003}]{shapelets}
{Refregier} A.,  2003, \mn@doi [\mnras] {10.1046/j.1365-8711.2003.05901.x},
  \href {https://ui.adsabs.harvard.edu/abs/2003MNRAS.338...35R} {338, 35}

\bibitem[\protect\citeauthoryear{Ricarte, Tremmel, Natarajan, Zimmer  \&
  Quinn}{Ricarte et~al.}{2021}]{wandering_bh}
Ricarte A.,  Tremmel M.,  Natarajan P.,  Zimmer C.,   Quinn T.,  2021, \mn@doi
  [\mnras] {10.1093/mnras/stab866}, 503, 6098

\bibitem[\protect\citeauthoryear{Sengül, Dvorkin, Ostdiek  \& Tsang}{Sengül
  et~al.}{2022}]{los_detection}
Sengül A.~C.,  Dvorkin C.,  Ostdiek B.,   Tsang A.,  2022, \mn@doi [\mnras]
  {10.1093/mnras/stac1967}, 515, 4391

\bibitem[\protect\citeauthoryear{{Sharon} et~al.,}{{Sharon}
  et~al.}{2022}]{Sharon2022}
{Sharon} K.,  et~al., 2022, \mn@doi [\apj] {10.3847/1538-4357/ac927a}, \href
  {https://ui.adsabs.harvard.edu/abs/2022ApJ...941..203S} {941, 203}

\bibitem[\protect\citeauthoryear{{Speagle}}{{Speagle}}{2020}]{2020MNRAS.493.3132S}
{Speagle} J.~S.,  2020, \mn@doi [\mnras] {10.1093/mnras/staa278}, \href
  {https://ui.adsabs.harvard.edu/abs/2020MNRAS.493.3132S} {493, 3132}

\bibitem[\protect\citeauthoryear{{Springel}, {Frenk}  \& {White}}{{Springel}
  et~al.}{2006}]{large_scales}
{Springel} V.,  {Frenk} C.~S.,   {White} S. D.~M.,  2006, \mn@doi [\nat]
  {10.1038/nature04805}, \href
  {https://ui.adsabs.harvard.edu/abs/2006Natur.440.1137S} {440, 1137}

\bibitem[\protect\citeauthoryear{Tremmel et~al.,}{Tremmel
  et~al.}{2018}]{Tremmel_2018}
Tremmel M.,  et~al., 2018, \mn@doi [\mnras] {10.1093/mnras/sty3336}, 483, 3336

\bibitem[\protect\citeauthoryear{Troxel et~al.,}{Troxel
  et~al.}{2018}]{cosmic_shear}
Troxel M.~A.,  et~al., 2018, \mn@doi [Phys. Rev. D]
  {10.1103/PhysRevD.98.043528}, 98, 043528

\bibitem[\protect\citeauthoryear{{Vega-Ferrero}, {Yepes}  \&
  {Gottl{\"o}ber}}{{Vega-Ferrero} et~al.}{2017}]{ellipticity_sim}
{Vega-Ferrero} J.,  {Yepes} G.,   {Gottl{\"o}ber} S.,  2017, \mn@doi [\mnras]
  {10.1093/mnras/stx282}, \href
  {https://ui.adsabs.harvard.edu/abs/2017MNRAS.467.3226V} {467, 3226}

\bibitem[\protect\citeauthoryear{{Yang}, {Birrer}  \& {Treu}}{{Yang}
  et~al.}{2020}]{linear_cluster_lensing}
{Yang} L.,  {Birrer} S.,   {Treu} T.,  2020, \mn@doi [\mnras]
  {10.1093/mnras/staa1649}, \href
  {https://ui.adsabs.harvard.edu/abs/2020MNRAS.496.2648Y} {496, 2648}

\bibitem[\protect\citeauthoryear{Şengül \& Dvorkin}{Şengül \&
  Dvorkin}{2022}]{effective_power_law}
Şengül A.~C.,  Dvorkin C.,  2022, \mn@doi [\mnras] {10.1093/mnras/stac2256},
  516, 336

\makeatother
\end{thebibliography}
\bsp

\label{lastpage}
\end{document}